\documentclass[prb,preprint,aps,showpacs,longbibliography,
lengthcheck]{revtex4-1}
\usepackage{graphicx}%
\usepackage{dcolumn}% Align table columns on decimal point

\begin{document}

\title{Nonlinearity exponents in lightly doped Conducting Polymers}

\author{D. Talukdar$^1$, U. N. Nandi$^2$, K. K. Bardhan$^1$}
\email{kamalk.bardhan@saha.ac.in}
\author{C. C. Bof Bufon$^3$} 
\altaffiliation{Institute for Integrative Nanosciences, IFW Dresden, Helmholtzstr. 20, 01069 Dresden, Germany}
\author{T. Heinzel$^3$, A. De$^1$}
\author{C.D. Mukherjee$^1$}

\affiliation{\\
$^1$Saha Institute of Nuclear Physics, I/AF Bidhannagar, Kolkata 700 064, India \\
$^2$Department of Physics, Scottish Church College, 1 $\&$ 3 Urquhart Square, Kolkata-700 006, India\\
$^3$Heinrich-Heine-Universit\"{a}t, Universit\"{a}tstrasse 1. 40225 Dusseldorf, Germany }

\date{\today}

\begin{abstract}
The \textit{I-V} characteristics of four conducting polymer systems like doped polypyrrole (PPy), poly 3,4 ethylene dioxythiophene (PEDOT), polydiacetylene (PDA) and polyaniline (PA) in as many physical forms have been investigated at different temperatures, quenched disorder and magnetic fields. Transport data clearly show the existence of a \textit{single} electric field scale for all systems. Based upon this observation, a phenomenological scaling analysis is performed, leading to extraction of a numerical value for a nonlinearity exponent called $x_M$ which serves to characterize a set of \textit{I-V} curves. The conductivity
starts deviating from an Ohmic value $\sigma_0$ above an onset electric field $F_o$ which scales according to $F_o \sim \sigma_0^{x_M}$. The electric field-dependent data are shown to be described by Glatzman-Matveev multi-step tunneling model [JETP 67, 1276 (1988)] in a near-perfect manner over nine orders of magnitude in conductivity and five order of magnitudes in electric field. Furthermore, $x_M$ is found to possess both positive and negative values lying between -1/2 and 3/4. There is no theory at present for this exponent. Some issues concerning applicability of the Glatzman-Matveev model are discussed.
\end{abstract}

\pacs{72.20.Ht,72.80.Le,05.70.Jk}

\maketitle

\section{Introduction}
Conducting polymers (CPs) such as doped polyacetylene (PA), polypyrrole (PPy), poly 3,4 ethylene dioxythiophene (PEDOT) and polyaniline(PANI) show a great variety of transport properties. In general, the electronic structure of $\pi$-conjugated pristine (undoped) conducting polymers originates from the s$p^2$$p_z$-hybridized wave functions of the carbon atoms in the repeat unit. Despite strong electron-phonon coupling in pristine CPs, an extraordinary large range of conductivities has been covered by doping. Conductivity $\sigma$ range from the highly insulating values $10^{-6}$ S/cm to highly metallic ones $10^5$ S/cm depending on the doping concentration \cite {epstein06}. Such a wide range of conductivities have made CPs useful in many applications such as wires, electromagnetic interference shields, anti-static coatings, conducting layers in active devices including organic- and polymer-based light emitting devices, photovoltaic devices, and field-effect transistors\cite {Wang,Kushto,Leefet}.    

The electronic transport properties of CPs are furthermore strongly influenced by the synthesis procedure and intrinsic disorder\cite {joo00, bufon05}. The conductivity of CPs originates from mobile charge carriers into the $\pi$-conjugated electronic orbitals which get filled by doping. At low doping densities, these charges self-localize to form solitons, polarons and bipolarons\cite {heeger88}. At higher doping levels, a transition to a metallic state is typically observed\cite {ishiguro92}. Quite often, disorder dominates the macroscopic properties,
thereby hiding or even eliminating the intrinsic delocalization
along the chains\cite {lee06}. As developed in Ref. \onlinecite {basescu87}, the intrachain transfer integrals are large as compared to the interchain transfer integrals, such that the transport takes place preferably along the chains. However, no matter how well the chains are ordered, the electrons finally have to hop between chains, and the related local resistances strongly influence or even dominate the macroscopic conduction. This leads to an increase in conductivity as the temperature is increased, i.e., an insulating behavior. Metallic behavior is observed
only at larger temperatures and only in some polymer systems\cite {kaiser01a}. Thermoelectric power(TEP) measurements as a function of temperature, as well as anomalous differences between optical and d.c. conductivity data, provide evidence\cite {kaiser01b} that CPs can be treated as being structurally heterogeneous, consisting of thin, metal-like fibrils, separated by potential barriers\cite {joo94}. This structural disorder is responsible for low electrical conductivities owing to low mobilities for most conducting polymers, even though in fully doped polymers, charge carrier concentrations can be as large as $10^{23}/{\rm{cm}}^3$, which is about four orders of magnitude higher than in highly doped inorganic semiconductors. However, highly doped CPs show an intrinsic metallic nature with traditional signatures like a Drude metallic response\cite {joo94, Kohlman95}, temperature independent Pauli susceptibility\cite {epstein87} and a linear dependence of the thermoelectric power on temperature \cite {kaiser89}. For the majority of CPs, the metallic state is strongly dependent on structural disorder, and only a small fraction of carriers remain delocalized at low temperatures with long scattering times. These delocalized electrons generate a weak temperature dependence of $\sigma$ down to millikelvin temperatures.

Polypyrrole has been particularly extensively investigated due to its many prospects for applications, like relatively high environmental stability, high conductivity, or the simplicity of preparation either by chemical or by electrochemical polymerization \cite {whang,nguyen99}. Polypyrrole is an amorphous conjugated polymer based on an aromatic ring and has a nondegenerate ground state \cite {street}. The polymer chains are intertwined and the fibrils are randomly oriented. Consequently, polypyrrole systems must be regarded as three-dimensional disordered systems with respect to their structure and morphology. Over the last few years, PEDOT has attracted a lot of interest because of high conductivity, optical transparency, easy processability and high stability\cite {kim03}. Inganas and co-workers\cite {pei94} have shown that PEDOT has a band gap of approximately 1.6 eV and can be cycled between the reduced and the oxidized state. It has been suggested that PEDOT has a lamellar-type structure built from ellipsoidal, conductive particles\cite {pingree08}. Even in highly conducting samples, one observes a mixture of insulating and metallic transport behavior. This has been explained in terms of a heterogeneous morphology\cite {kaiser01b}.

Conducting polymers are known to undergo a metal-insulator transition as a function of doping\cite {yoon94}. Samples at the insulating side of the transition are of interest here. Conduction data\cite {amitava94,sanjib07, ribo98,aleshin04,bufon07} in CPs at low fields in this regime are usually discussed within the framework of the standard Mott variable-range hopping (M-VRH) model\cite {mott69} and its modification (ES-VRH) by Efros and Shklovskii\cite {efros75} in presence of Coulomb interactions at low temperatures. In any VRH model, an electron near the Fermi level hops between localized states by absorbing phonons and the conductivity $\sigma$ is assumed to follow from path optimization of hopping distance and difference in energy levels involved:
\begin{equation}
\sigma(T) = \sigma_m {\rm{exp}} \left [-\left ({T_0 \over T}\right )^m \right ].  \label{eq:mott}
\end{equation}
Here $\sigma_m$ is the conductivity pre-factor and \textit{T} the temperature. In the M-VRH model, the exponent \textit{m} is given by $m = 1/(d+1)$ where $d$ is the dimensionality. Thus, \textit{m} is 1/4, 1/3 and 1/2 in three, two and one dimension respectively. The characteristic temperature $T_o$ is given by $T_o=4\pi /3N(E_F)k_B a^3 $ where $N(E_F)$ is the density of states at the Fermi level, $k_B$ the Boltzman constant and \textit{a} the localization length. In the ES-VRH model, interactions among electrons were shown to open up a coulomb gap in the density of states at the Fermi level. This leads to $m=1/2$ independent of the dimension and
\begin{equation}
T_o={2.8e^2 \over 4\pi \epsilon_0 \epsilon k_B a},    \label{eq:To}
\end{equation}
where $\epsilon$ is the dielectric constant. The standard VRH model seems to describe well three dimensional systems like powdered FeCl$_3$-doped PPy\cite {amitava94}, PEDOT at higher temperature\cite {sanjib07} yielding \textit{m}=1/4. PTS-doped PPy films\cite {ribo98} are reported to yield \textit{m}=1/3 suggesting two-dimensional nature. The exponent in single crystals of PTS-doped polydiacetylene (PDA)\cite{aleshin04} is found to be $\sim$0.65-0.70 apparently demonstrating quasi-one dimensional nature of the system. On the other hand, some samples like PEDOT at low temperature\cite {sanjib07}, PPy films\cite {bufon07} tend to follow ES-VRH particularly at low temperature with \textit{m}=1/2. Apart from the VRH model, another model\cite {sheng78} called fluctuation-induced tunneling (FIT) model also has been used to describe electrical conduction in CP's, especially in quasi one-dimensional structures such as nanotubes or nanofibres. This model characterizes electrons transfer across the insulating barriers between conducting regions and hence, can be appropriate for describing the transport properties of CPs as individual polymer fibrils have a mixture of both crystalline and non-crystalline insulating regions. In this model, the temperature dependence of $\sigma$ is given by 
\begin{equation}
\sigma (T)=\sigma_h {\rm{exp}} \left [-\frac {T_b}{T_s + T} \right ], \label{eq:fit}
\end{equation}
where $\sigma_h$ is a temperature-independent prefactor, $T_b$ represents the typical magnitude of tunneling barriers, and the ratio $T_s/T_b$ determines the extent of reduction in conductivity at low temperatures. The model, originally developed to explain the temperature dependence of conductance in C-PVC composites\cite {sheng78}, was subsequently applied to doped polyacetylene\cite {sheng80}, isolated nanotubes or nanofibres \cite {kaiser03,kaiser04a}. It is known that the latter often exhibit conduction properties different from those of its assemblies. The two models differentiate themselves in a fundamental manner by their behavior as \textit{T} approaches zero - the VRH conductivity goes to zero even as the FIT one remains finite. As seen later, this will have a bearing on the nonlinear behavior. Note that both the models in special cases (i.e. \textit{m}=1 in VRH or $T_s =0$ in FIT) can represent an activated process.
In rest of the paper, we will use the symbol $\sigma$ for conductivity and $\Sigma$ for conductance. The latter here is simply taken as the chordal conductance, \textit{I/V} unless otherwise mentioned.
 
Non-Ohmic or nonlinear conduction is a very common feature of disordered systems in general, and conducting polymers in particular. Onset of nonlinearity in these systems often takes place upon application of only few volts of bias across samples in laboratories. In bulk systems, this is rendered possible by the presence of microscopic inhomogeneties which lead to very large local fields. Now, with advent of low dimensional materials such as nanotubes, nanofibres, nanowires or quantum dots, the generation of large fields with moderate bias applied across small lengths are quite common. This provides added incentive to study and understand the phenomena of nonlinear conduction in CPs. Nonlinear transport data in conducting polymers \cite {nguyen99, park02, jgpark2, jgpark3, kaiser04a, kaiser05, aleshin04, bufon07, zhi09} are either presented in form of \textit{I-V} curves or equivalently, in form of \textit{$\Sigma$-V} curves, measured as a function of temperature. However, there is hardly any systematic analyses of the data mainly because theoretical understanding \cite{hill71,shklov73,pollak76, shklov76, sheng78, park02, matveev88} is far from complete. For example, the models of variable range hopping under field \cite{hill71,pollak76,shklov73, shklov76} have expressions for the field-dependent conductivity in two field limits - low or moderate and high - as a function of temperature. Glazman and Matveev (GM)\cite {matveev88} have evaluated a model involving multi-step tunneling across thin disordered regions also in two limits, $eV \gg k_B T$ and $eV \ll k_B T$. In the first limit, one has at any temperature a full expression for $\Sigma(V)$ containing parameters without having any temperature dependence.

In this paper, we present electrical transport measurements covering both linear and non-linear regimes in moderately doped PPy in forms of pellets as well as films and in a pellet of doped PEDOT over a wide range of temperatures. In addition, some published data from literature have been also processed for comparison. For analysis of our data, we depart from methods used in existing literatures and adopt a newly suggested methodology\cite {taluk10} based upon observation of existence of a field scale in many disordered samples. This method of scaling analysis of a set of \textit{I-V} curves yields an exponent $x_T$, called nonlinearity exponent, that  characterizes the nonlinear data and is believed to reflect the underlying field-dependent conduction mechanism. We apply this scaling approach to various conducting polymer systems, examine its validity and extract the nonlinearity exponents. In the next section (section II), the method of scaling analysis is described in detail. Furthermore, two models which are extensions of those earlier considered at low fields, namely, VRH and FIT models are also reviewed along with corresponding experimental results reported in the literature. The GM-model\cite{matveev88} was originally developed for thin amorphous films whose dimensions along the hopping direction (i.e. thickness) lie between the localization lengths and hopping lengths and has never been used before to explain nonlinear data in CPs. It is found rather surprisingly to be very successful in explaining the \textit{I-V} characteristics in CPs. These models are critically discussed in particular context of the general requirements of scaling. Experimental details are given in section III. Data are presented and analyzed mainly by adapting the GM-expression as a scaling function in section IV. Results so obtained are then discussed in section V. Finally, conclusions are given in section VI.

\section{Survey of models of nonlinear conduction: Theoretical and Phenomenological}

As the form of disorder can be of huge variety, so could be the mechanisms causing deviations from Ohmic behavior. Here we will concern ourselves with the type of non-Ohmic phenomena that arise in bulk due to interplay of disorder and various fundamental processes such as localization, tunneling, hopping etc. Thus, we exclude from our purview cases like space charge related conduction in organic solids\cite {} and joule heating, that also occur in bulk. The joule heating is, of course, present in any conductor, whether ordered or disordered, and increases with passing current. In case of positive temperature coefficient of resistance (as in metals), the Joule heating leads to an increase in resistance whereas in case of negative temperature coefficient of resistance (as in systems under consideration), it leads to a decrease in resistance. Since both the applied bias and Joule heating can lead to decrease in sample resistance, care must be taken during experiments to ensure that effects of joule heating is insignificant or taken it into account in the interpretation of data.

A study of field-dependent conduction is generally expected to bring out subtleties or additional processes that are either not present or insignificant in the linear conduction. Consider, for example, composites which are random mixtures of conductors and insulators above the percolation threshold. For small bias, conduction takes place only through the backbone and is Ohmic. As the bias is increased, tunneling across the thin insulating layer between the ends of the conductor chains dangling off the backbone are believed to lead to increased conductance\cite {kkb97}. Furthermore, a theoretical treatment of non-ohmic conduction may lead to self-consistent relations among parameters that are also  involving in Ohmic conduction. This is illustrated in case of VRH under field (see below) where physical parameters like localization length may be involved in more than one experimentally measurable quantities. In view of the inadequate theoretical development it should be of much utility to know general properties of nonlinear conduction. One can then check whether any specific theoretical model prediction is consistent with general requirements or not. One general property is that the conductance $\Sigma$ always increases with field \textit{F}, at least at small fields. This property is obvious since the application of a field results in lowering barrier heights, thus decreasing resistance to conduction. For a proper description, one defines a characteristic field or a field scale $F_o$ such that $\Sigma(F) \ge \Sigma(0) $ for $F \ge F_o$. This corresponds to the fact that a sample remains Ohmic at small fields and starts deviating from the Ohmic behavior as the field is increased beyond $F_o$. The scale may also be given formally by the following ($\Delta \sigma(F) = \sigma(F) - \sigma(0)$):
\begin{equation}
\Delta \sigma(F_o) \sim \sigma(0).
\label{eq:deffo}
\end{equation}
\textit{I-V} curves are generally measured as a function of some physical parameter upon which $\sigma(0)$ depends. Temperature is the most commonly used parameter. Others like magnetic field, pressure are also valid parameters. An interesting but less studied situation is when \textit{I-V} data are gathered simply as a function of quenched disorder as in composite samples\cite {kkb97}. Naturally, $F_o$ is expected to be a function of the parameter.

\subsubsection{Scaling analysis: Nonlinearity exponent}

Let us consider, for definiteness, cases where measurements are carried out at different temperatures. Recently, a generalized approach based upon existence of a \textit{single} field scale in a given disordered sample has been adopted, leading to characterizing a set of \textit{I-V} curves by a single number $x_T$, called nonlinearity exponent \cite{taluk10}. In this approach, the conductivity $\sigma (T,F)$ is given by the scaling relation:
\begin{equation}
{\sigma(T,F) \over \sigma(T,0)} = g \left ({F \over F_o} \right ),
\label{eq:scaling}
\end{equation}
where \textit{g} is a scaling function and the field scale $F_o(T)$ at each temperature is given by the phenomenological relation
\begin{equation}
F_o (T) = A_T \; {\sigma_o}^{x_T},
\label{eq:fscale}
\end{equation}
where $\sigma_o(T) = \sigma(T,0)$ is the conductivity at zero bias at temperature $T$ and $A_T$ is a constant whose value depends upon the criterion that fixes the scale $F_o$. For  $F \le F_o$, the scaling function $\textit{g} \approx 1$ corresponding to the fact that the conductance increases very little from the zero field value $\sigma_o$. At larger fields $F > F_o$, $\textit{g} > 1$. Thus, the field $F_o$ can be called a crossover or onset field such that it separates the linear regime from the non-linear regime along the field axis. Note that temperature does not enter \textit{explicitly} in Eq. (\ref{eq:fscale}) but does so through the temperature-dependent $\sigma_o(T)$. Thus, according to Eq. (\ref{eq:fscale}) the field scale for nonlinearity is determined solely by the linear conductivity $\sigma_o$. This is indeed a very significant relation in that the same relation holds good even when $\sigma_o$ is changed by some other variable like magnetic field or simply quenched disorder (see Eq. \ref{eq:pfscomp}). Presently, a theoretical understanding of this result is lacking but diverse disordered systems with localized states\cite {taluk10} including amorphous and doped semiconductors, composites at low temperatures have been found to obey such scaling, all with \textit{positive} exponents $x_T$.

At large fields $F \gg F_o$, \textit{I-V} curves often tend to become \textit{independent} of temperature. From Eq. (\ref{eq:scaling}) this follows immediately if $g(x) \sim x^{1/x_T}$ at large \textit{x} for $x_T \ge 0$. Thus, at large fields the conductivity varies as a power-law with an exponent $z_T$:
\begin{equation}
\sigma(T,F) \sim F^{z_T},~~z_T=1/x_T~ \textrm{for}~ x_T > 0.
\label{eq:large}
\end{equation}
This prediction provides a self-consistency check for the exponent $x_T$ as it may be determined using two independent methods: one from a set of \textit{I-V} curves at low fields and the other from a single \textit{I-V} curve at high fields. This has been also amply verified\cite {taluk10} in systems mentioned earlier. It is also clear from this that in case of negative $x_T$ one can not expect temperature-independence of conductivity at large field although the scaling itself can not specify any particular functional form.

Composite systems at room temperature are presently the only disordered ones where the nonlinearity exponent has been explained\cite {gefen86, cbb91} although they are not, strictly speaking, comparable to the systems under study in this paper. Firstly, conduction in composites (at least) at room temperature is not known to take place through localized states and is rather metallic since resistance increases with temperature. Secondly, it is not clear how results in a percolating network\cite {stauffer} (which composites are supposed to be prototypes of) could be directly applied to disordered systems such as CPs since a hopping network is considered to be a percolating network always at criticality\cite {efrosbook} with infinite correlation length.  Nevertheless, it may be instructive to review non-Ohmic conduction in these systems. In this case, \textit{I-V} curves have been measured at a \textit{fixed} temperature (i.e. room temperature) in samples with various degree of disorder characterized by the parameter \textit{p}, fraction of conductors in a sample. Near the percolation threshold $p_c$ (i.e. $p \ge p_c$), \textit{I-V}'s turn out to be nonlinear even at room temperature with conductance increasing with the applied bias. As mentioned earlier, the increase in conductance is due to opening up of new channels of conductions - from tunneling across closely spaced tips of pairs of branches of conductors dangling off the backbone. The bias scale was found to vary as the inverse of the correlation length $\xi$:
\begin{equation}
V_o (p) \le {\xi}^{-1} \sim {\Sigma_o}^{\nu /t}, \label{eq:pfscomp}
\end{equation}
so that the nonlinearity exponent $x_p$ is given by $x_p \le \nu / t $ where $\nu$ and \textit{t} are the correlation and conductivity exponents respectively and $\Sigma_o(p)=\Sigma(p,V=0) \sim {(p-p_c)}^t$. The subscript in the exponent denotes the parameter that is varied to change $\Sigma_o$. $\nu / t$ is about 0.45 in three dimension and consistent with the values of $x_p$ obtained experimentally in discontinuous gold film\cite{gefen86} and carbon-wax\cite {cbb91}. In a sample of the same carbon-wax system with a given $p$ (i.e. \textit{fixed} disorder), \textit{I-V} curves were measured at different temperatures \cite {nb2}. The onset voltage $V_o(T)$ scaled as before with the linear conductance $\Sigma_0(T)=\Sigma(T,V=0)$ as
\begin{equation}
V_o(T) \sim {\Sigma_o}^{x_T}.        \label{eq:Tfscomp}
\end{equation}
Interestingly, $x_T$ was found to have the same value as $x_p$ defined in Eq. (\ref {eq:pfscomp}). 

It is seen from Eqs. (\ref{eq:fscale}) and (\ref{eq:large}) that the nonlinearity in a system can be characterized by two quantities: the field scale $F_o$ that determines the onset and the nonlinearity exponent $x_M$ that determines the degree (i.e. steepness of increase of the conductivity with field). Both the scaling function \textit{g} and the nonlinearity exponent $x_T$ are obviously determined by the details of the conduction mechanism under study. In principle, the problem of non-Ohmic conduction is solved if an expression for \textit{I(M,V)} is found where the \textit{I-V} data are taken at different values of the parameter \textit{M}. Let us now consider several models and see how they conform to general scaling formulation.

\subsubsection{Variable range hopping under field}

In disordered systems, an increasing electric field aligns an increasing number of empty and accessible states to the occupied states allowing charge carriers to move via phonon-assisted tunneling or hopping transitions. With decreasing temperature the mean hopping length $R_h$ grows as  
\begin {equation}
R_h= (a/2)(T_o/T)^m,
\label{eq:rh}
\end {equation}
where $T_0$ and \textit{m} are same as in Eq. (\ref{eq:mott}). With higher electric fields the energy $eFR_h$ gained by an electron may become comparable to $k_B T$ and lead to deviation from Ohmic behavior. Theories \cite {hill71,pollak76,shklov76} predict two characteristic fields $F_l$ and $F_u$ such that the non-Ohmic conductivity at intermediate fields  $F < k_B T/e a$ is given by
\begin{equation}
\sigma(T,F) = \sigma(T,0) \; {\rm{exp}} \! \left (\frac {eFL}{k_B T} \right ),
\label{eq:lowF}
\end{equation}
where $\sigma(T,0)$ is given by Eq. (\ref{eq:mott}) and \textit{L} is a length related to the hopping length $R_h$. Clearly, $F_l = k_B T/eL$ could be identified with the onset field scale $F_o$ discussed above. \textit{L} is proportional to $R_h^{\mu}$ where $\mu$ is either 1 (Refs. \onlinecite {hill71,pollak76}) or 2 (Ref. \onlinecite {shklov76}). Thus, it follows from Eqs. (\ref{eq:lowF}) and (\ref{eq:rh}) that $F_o$ varies with \textit{T} as a power-law:
\begin {equation}
F_o \approx F_l \sim T^\alpha /a {T_o}^{\alpha -1},~~~ \alpha = 1+m\mu,
\label{eq:FT}
\end {equation}
with the same \textit{m} as in Eq. (\ref{eq:mott}). According to Eq. (\ref {eq:FT}), the exponent $\alpha$ is always positive and greater than 1. In the literature however, authors traditionally focus on the temperature dependence of $L \sim T^{-m\mu}$ rather than the field scale. The compliance of this with experimental results\cite {ladieu00} in amorphous and doped semiconductors is rather poor. In many cases the power-law (i.e, a straight line in a log-log plot of $F_o$ vs. \textit{T}) is not observed. When the plot appears like a straight line, the exponent often turns out to be randomly different from the expected value in Eq. (\ref {eq:FT}) albeit with few exceptions. In the limit of large fields $F \ge k_B T/ea$ theories\cite {shklov73,pollak76} agree with 'activationless' hopping at least qualitatively. In such situation the energy gained by an electron is large enough to hopp without any need to absorb any phonon. As a consequence, the conductivity becomes independent of temperature and is given by
\begin{equation}
\sigma \sim  {\rm{exp}} \left [- \left ( {F_u \over F} \right )^m \right ],~~~~F_u = a_1 \frac {k_B T_o}{ea},
\label{eq:highF}
\end{equation}
with the same \textit{m} and $T_o$ as in Eq. (\ref{eq:mott}). $a_1$ is a numerical constant equal to unity\cite {shklov73} when \textit{m}=1/4. Activationless hopping has been observed by many authors\cite {ladieu00} irrespective of the value of \textit{m}. However, inexplicably the particular expression of high field conductivity in Eq. (\ref{eq:highF}) has been found so far \textit{only} in systems with \textit{m}=1/2. On the contrary, in several systems\cite {morgan71, *servini69} with \textit{m}=1/4, the field dependence is well described, albeit empirically, by a power-law $\sigma \sim F^z$ reminiscent of Eq. (\ref{eq:large}). As mentioned earlier, Eqs. (\ref{eq:highF}) and (\ref{eq:To}) illustrate the possibility of having self-consistent relations involving the same parameters such as \textit{a}, $T_o$. This has led to quantitative disagreement\cite {bufon07,yu04} in that experimental values of $F_u$ turn out to be always greater than the calculated ones.

In presence of coulomb interactions with \textit{m}=1/2, some authors\cite {dvure88,yu04,bufon07} have suggested that the field-dependent conductivity can be written in the following manner:
\begin{equation}
\nonumber
\sigma(F,T) \approx {\rm {exp}} \left ( -\frac{2R_h}{a}- a_1 \frac{a T_0}  {4R_hT} + \frac{eFR_h}{k_BT} \right ). 
\end{equation}
In the activationless regime the last two terms in the exponent cancel each other and Eq. (\ref{eq:highF}) is recovered. But there is no unanimity in the value of $a_1$ which can be 1/2\cite {yu04}, 1\cite {dvure88} or 1.44\cite {bufon07}. More significantly, it is to be noted that while the above expression at least in intermediate fields is compatible with the scaling (Eq. \ref{eq:scaling}), the large field conductivity as given by Eq. (\ref{eq:highF}) is not.

\subsubsection{Fluctuation-induced tunneling under field}

Kaiser et al.\cite {kaiser04a,kaiser05} extended the usual FIT model\cite {sheng78} at small fields by including the backflow current, the tunneling near the top of the barrier and thermal activation over the barrier to obtain the following phenomenological expression for conductivity:
\begin{equation}
\sigma= \frac{\sigma_0 \; {\rm{exp}}(V/V_o)} {1+ h \; [{\rm{exp}} (V/V_o)-1]},        \label{eq:kaiser}
\end{equation}
where $\sigma_0$ is given by Eq. (\ref{eq:fit}) and the parameter $h = \sigma_0/\sigma_{\infty}$ ($h \le 1$), $\sigma_{\infty}$ being the value of $\sigma$ at large voltages \textit{V}. $V_o$ is a voltage scale factor. At small bias $V \le V_o$, the conductivity increases exponentially with \textit{V} but at large bias $V \gg V_o$, the conductivity increases to the saturation value $\sigma_ {\infty} $. The quantities $V_o$ and $\sigma_ {\infty} $ have not been related to parameters of the model and their temperature dependence remain unknown. The above expression is claimed to give a good description of the nonlinear \textit{I-V} characteristics at different temperatures in quasi-1D systems like PA nanofibre, single wall carbon nanotube network \cite {kaiser04a}. However, Yin et al.\cite {zhi09} reported systematic deviations between the \textit{I-V} data and the fitting curves at low voltages in several quasi-one dimensional systems like individual PPy nanotube, PEDOT nanowire, PA nanotube, and CdS nanopore although overall fits appeared reasonable. Moreover, \textit{h} was found to have a substantial dependence upon temperature, increasing by as much as two orders of magnitude with temperature in case of PPy nanotube. This is contrary to expectations since $\sigma_ {\infty} $ is expected to be independent of temperature at large bias. There have been also several instances\cite {kaiser05} where $h \approx 0$ indicating that $\sigma_ {\infty} $ is probably very large. As discussed above, it is the variation of $V_o$ with temperature or any other variable that is of interest here.   

\subsubsection{Multi-step tunneling}

Considering the process of multi-step indirect tunneling via \textit{n}-localized states in a disordered system, Glazman-Matveev (GM) proposed the following expression\cite {matveev88} for the conductance through a amorphous semiconductor thin film (i.e. tunnel barrier) of thickness \textit{w} under bias \textit{V} ($eV \gg k_BT$ and $p_n = n- 2/(n+1)$): 
\begin{eqnarray}
\nonumber
\Sigma & = & {\Sigma_d}+\sum^{n}_{1}{\Sigma_n}{V^{p_n}} \\  \nonumber
       & = & {\Sigma_0}+{\Sigma_2} V^{1.33} + {\Sigma_3} V^{2.5} + {\Sigma_4} V^{3.6} + \\
       & & {\Sigma_5} V^{4.67} + {\Sigma_6} V^{5.71} +...,        \label{eq:gms}
\end{eqnarray}
where ${\Sigma_0 = \Sigma_d+\Sigma_1}$. $\Sigma_d$ accounts for the direct tunneling and $\Sigma_1$ for the elastic resonant tunneling via one localized state. Each term ($n>1$) in the series arises out of rare events when a number of localized states happen to be arranged physically as well as energetically in such a way that an electron can traverse a sample length via multi-step inelastic tunneling. Each term may be thought to constitute a separate channel of conduction involving a definite number of localized states. Thus, the macroscopic nonlinearity in this GM-model results from two contributions - primarily,  appearance of a new channel with increasing bias and secondarily, nonlinearity of each such channel. This multi-step tunneling model has been widely invoked to explain relevant data in various tunnel junctions\cite {beasley95}, manganites \cite {gross00,markovich}. Interestingly, the manganites samples were not necessarily in the form of junctions. Nevertheless, it was thought fit to apply GM-model because of possibilities of tunneling across grain boundaries, or insulating barriers, separating metallic phases within samples. Considering the similar structure of conducting polymers i.e. the polymer fibrils consisting of quasi-metallic lengths with intervening insulating regions, CPs are expected to be also candidates for application of the GM-model. It is shown below how a couple of assumptions about the coefficients $\Sigma_n$ make the GM-expression (Eq. \ref{eq:gms}) not only compatible with the general scaling formulation (Eq. \ref {eq:fscale}) but also yield the same relation between the nonlinearity exponent $x_T$ and the bias exponent $z_T$ as in Eq. (\ref {eq:large}).

\textit{Scaling}: While the \textit{I-V} data at different temperatures have been fitted well by Eq. (\ref{eq:gms}) with few nonlinear terms\cite {gross00,markovich}, no attempt was made to analyze systematically the temperature variation of the coefficients $\Sigma_n$ in Eq. (\ref{eq:gms}) . Note that Eq. (\ref{eq:gms}) as written is not compatible with the scaling (Eq. \ref{eq:scaling}). One way to achieve this is to assume that for $n \ge 2$, the coefficients satisfy the following
\begin{equation}
\Sigma_n = c_n \; \Sigma_0 {V_0}^{-p_n},
\label{eq:gmcoeff}
\end{equation}
so that Eq. (\ref{eq:gms}) can be written as
\begin{eqnarray}
\nonumber
\Sigma \over \Sigma_0 & = & 1 + c_2 {\left ( V \over V_0 \right )}^{1.33} + c_3 {\left ( V \over V_0 \right )}^{2.5} + c_4 {\left ( V \over V_0 \right )}^{3.6} + \\
       & & c_5 {\left ( V \over V_0 \right )}^{4.67} + c_6 {\left ( V \over V_0 \right )}^{5.71} +...        \label{eq:scgms}
\end{eqnarray}
The above expression explicitly implies the existence of a single bias scale $V_o$ as in Eq. (\ref{eq:scaling}). The right hand side of the above equation is thus really a scaling function as defined in Eq. (\ref {eq:scaling}). The same holds if conductivity and field are used in place of conductance and bias. It is seen from Eq. (\ref{eq:gmcoeff}) that the constants $c_n$ are determined once the coefficients $\Sigma_n$ and scale $V_o$ are known. The expression (Eq. \ref{eq:scgms}) suggests that experimental \textit{$\Sigma$-V} curves at different \textit{T}'s may be collapsed into a single curve by suitable choices of $\Sigma_o$ and bias scale $V_o$ at each temperature. Note that at $V = V_o$, the coefficients $c_n$'s satisfy the following relation:
\begin{equation}
{\Sigma(V_o) \over \Sigma_0} =  1 + \sum^{n}_{2}{c_n}.
\label{eq:cns}
\end{equation}
If, for example, $ \Sigma(V_o) = 2\Sigma_o$, we have $ \sum^{n}_{2}{c_n} =1$.

\textit{Nonlinearity exponent} $x_T$: The GM model does not give any guidance on the temperature dependence of the coefficients $\Sigma_n$ under the condition $eV \gg k_BT$ for which Eq. (\ref{eq:gms}) is valid. Therefore, one needs to fall back on the empirical relation (Eq. \ref {eq:fscale}). Alternatively, if the coefficients $\Sigma_n$ are known from fittings one can then find $V_o$ in the following manner. Let $n=n_o$ be the lowest channel with nonzero $\Sigma_{n_o}$. From Eqs. (\ref {eq:deffo}) and (\ref {eq:gms}) the onset bias $V_o$ is given by $\Sigma_{n_o} {V_o}^ {p_{n_o}} \sim {\Sigma_o}$ which leads to the expression for $V_o$ as 
\begin{equation}
V_o \sim \left ( {\Sigma_0 \over {\Sigma_{n_o}}} \right )^{1 /p_{n_o}}. 	\label{eq:g2}
\end{equation}
This is consistent with the more general assumption (Eq. \ref{eq:gmcoeff}). In absence of any theoretical guidance, we make here a second assumption that $\Sigma_n$ follows a power law behavior with ${\Sigma_0}$ as given by 
\begin{equation}
\Sigma_n \sim {\Sigma_0}^{y_n}, \label{eq:yn}
\end{equation}
where $y_n$ is an exponent. Incorporating this result in Eq. (\ref{eq:g2}) and comparing with Eq. (\ref{eq:fscale}), one has 
\begin{eqnarray}
x_T ={ {1-y_{n_o}} \over {p_{n_o}} } = { {1-y_n} \over {p_n} },  \label{eq:ne2}
\end{eqnarray}
for any $n \ge 2$. The second equality on the right hand side of the above equation follows from Eqs. (\ref{eq:gmcoeff}) and (\ref{eq:yn}). Notice that while the left hand side of Eq. (\ref{eq:ne2}) is a constant, the right hand side carries the index \textit{n}. Thus, if $n_o=2$, then $x_T = {3 \over 4}(1-y_2)$. If $n_o=3$, then $x_T = {2 \over 5}(1-y_3)$ and so on. All three phenomenological relations Eqs. (\ref{eq:fscale}), (\ref{eq:gmcoeff}), and (\ref{eq:yn}), only two of which are independent, underline a basic assumption that nonlinear scales are determined by the corresponding linear conductivity. Eq. (\ref{eq:ne2}), a somewhat generalized version of the relation considered earlier by Chakrabarty et al.\cite {cbb91}, is remarkable for several conclusions that immediately follow from it:

i) $x_T$ is 0 when $y_{n_o}=1$. It is positive or negative depending upon whether $y_{n_o}$ is less or greater than 1. Furthermore, its maximum  possible value is $1/p_2$ or 0.75 as $p_n$ has the lowest value for \textit{n}=2. Thus, we have $-1/2 \le x_T \le 3/4$. The lower limit of -1/2 is obtained from general arguments\cite {taluk10}.

ii) Since the left hand side of Eq. (\ref{eq:ne2}) is independent of \textit{n}, and $p_n$ increases with \textit{n}, it follows that for $x_T > 0$, $y_n$ must decrease with \textit{n} and for $x_T < 0$, $y_n$ must increase with \textit{n}.
\begin{eqnarray}
\nonumber
1 \geq y_2 \geq y_3 \geq .. \geq y_n,~~~x_T \geq 0 \\
1 < y_2 < y_3 < .. < y_n,~~~x_T < 0 
\label {eq:y1yn} 
\end{eqnarray}
Thus for $x_T < 0$, we have $y_n = -x_T p_n$ for $y_n \gg 1$ at large \textit{n}. If any $\Sigma_n$ in the GM expression is zero, the corresponding $y_n$ is obviously excluded from the above.

iii) For $x_T > 0$, the above naturally raises question about the limit of $y_n$ for large \textit{n}. Considering the fact that the conductance of a hopping system at large bias may tend to become independent of \textit{T} or in other words, $\Sigma_0$, the lower limit of $y_n$ may be taken as zero i.e. as $n \rightarrow \infty$, $y_n \rightarrow 0$. Note that $y_n$ is always a positive quantity since the coefficient $\Sigma_n$ is predicted to be a decreasing function of \textit{n}\cite {matveev88}.

iv) Since $p_n$ increases with \textit{n}, Eq. (\ref{eq:ne2}) requires that for $x_T > 0$, the series (Eq. \ref{eq:gms}) must terminate at $n = n_L$ where the highest channel $n_L$ is given by $p_{n_L-1} < 1/x_T \le p_{n_L}$.

\textit{Large bias exponent} $z_T$: If the field-dependent conductance of a system continues to be described by Eq. (\ref{eq:gms}) (e.g. there is no appearance of negative differential conductance) it follows that at large bias or field the conductance varies with bias as a power-law: $\Sigma \approx \Sigma_{n_L} V^{p_{n_L}}$ where $n_L$ is the index of the highest allowed channel (see the point (iv) above). Such a power-law variation is also predicted by the scaling consideration (Eq. \ref{eq:large}) for $x_T >0$ with $z_T = {p_{n_L}}$. Now, from Eq. (\ref {eq:ne2}), we obtain $x_T = ({1-y_{n_L}}) / p_{n_L} $ so that $z_T = ({1-y_{n_L}})/x_T  \cong 1/x_T$ following the point (iii). Thus, we obtain $z_T \cong 1/x_T$ as follows from general scaling consideration (Eq. \ref{eq:large}).

\begin{table}[b]
\caption{\label{tab:table1}%
Sample growth conditions and parameters (Eq. \ref{eq:mott}).}
\begin{ruledtabular}
\begin{tabular}{cccccr}

\textrm{System}&
\textrm{Sample}&
\textrm{Oxidant} &
\textrm{$\sigma_o$(300K)} &
\textit{m} &
\textrm{$T_o$} \\
&
\textrm{No.}&
&
\textrm{S/cm} &
&
\textrm{K} \\

\colrule
PPy (powder)    &1 & FeCl$_3$  &5.7x$10^{-4}$ & 1/4 & 1.7x$10^6$  \\ 
PEDOT	(powder)	&2 & FeCl$_3$       &2.6x$10^{-2}$ & 1/2 &1798  \\ 
PPy (film)	  &3 & HCl:H$_2$O$_2$ &10 & 1/2 &2340 \\ 
PPy (film)  	&4 & HCl:H$_2$O$_2$ &12 & 1/2 & 950   \\
PPy (film)  	&5 & HCl:H$_2$O$_2$ &8.5 & 1/2 &1600  \\
 				
\end{tabular}
\end{ruledtabular}
\end{table}

\section{Experimental}

\begin{figure}
\includegraphics[width=7cm]{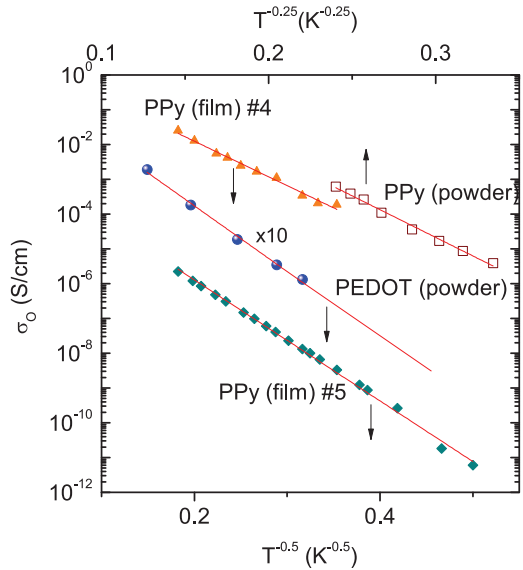}
\caption{(Color online) Variation of Ohmic conductivity $\sigma_0$ vs. $T^{-m}$ with $m=1/4, 1/2$ for four conducting polymer samples as shown. For clarity, the PEDOT data have been shifted upwards by a factor as shown. Solid lines are linear fits to the data.}
\label{fig.1}
\end{figure}

The three systems used in this study were doped PPy in powder and film forms, and doped PEDOT in powder form. All together five samples (see Table I) have been studied. All samples were synthesized using chemical polymerization. To aqueous solution of pyrrole monomer (0.05 M), 100 ml of aqueous FeCl$_3$ (0.1 M) solution was added dropwise leading to almost instant polymerization. Black precipitate of polypyrrole was separated from the solution and the resulting powder was then vacuum dried and pressed into disk shaped pellets, 8 mm in diameter and 1 mm in thickness (sample 1). Further details can be found in Ref. \onlinecite {amitava94}. A similar procedure was followed for synthesis of doped PEDOT (sample 2) using monomer ethylenedioxythiophene (EDOT) dissolved in water with 0.2M dodecylbenzenesulphonic acid. Subsequently, an aqueous solution of the oxidizing agent FeCl$_3$ was added dropwise on constant stirring under inert atmosphere. EDOT:FeCl$_3$ molar ratio was 1.167:1. After overnight reaction, the precipitate was thoroughly washed with ethanol, vacuum dried, and pelletized in form of disks of same dimensions as above. PPy films (samples 3-5) were synthesized from pyrrole vapor in solutions of HCl:H$_2$O$_2$=1:1000 while the solvent was cooled to 282, 283 and 300 K respectively. Further details can be found in Ref. \onlinecite {bufon07} (the sample 3 here is the same as the sample A in the latter). Thicknesses were 70, 77 and 60 nm respectively.

Table I shows typical room temperature conductivities of the samples. For transport measurements, thin copper wires were attached to both ends of the disks using silver paint whereas leads were connected to pre-patterned Pt electrodes on the substrate containing PPy film\cite {bufon07}. Two and four probe dc transport measurements gave similar results indicating negligible contact resistances in samples. Low temperature measurements were done in liquid helium cryostats in the temperature range 2.1-300 K. The PPy and PEDOT samples were placed on sapphire substrates with Apeizon N-grease and data were taken under the constant current condition. In case of PPy films, data were taken under the constant voltage condition. Temperature was stabilized to better than $\pm50$ mK for the \textit{I-V} measurements. Maximum current levels were kept low to minimize joule heating in the samples. For example, the maximum current through the PEDOT sample at 26 K was limited to 0.22 mA as an increased current led to instability due to joule heating. However this behavior was completely reversible and sample resistance returned to the initial value after removal of current. 

\section{Results}

Fig. 1 shows Ohmic conductivities $\sigma_o$ vs. $T^{-m}$ for four samples - one each of PPy (powder) and PEDOT, two of PPy (film). Similar data for sample 3 have been already presented in Ref. \onlinecite{bufon07}. Excellent linearity in the data shows that all the samples obey variable range hopping conduction in the range of temperatures measured. While \textit{m} is $1/4$ in PPy (powder), it is $1/2$ in all samples of other two systems. There is a transition at 30 K in PPy films from ES-VRH mechanism to activated process. A similar transition from ES-VRH to M-VRH also takes place in PEDOT at about 38 K. All the samples were in insulating regimes. $T_0$'s obtained from slopes of the fitted lines are of the order of $10^3$ K and are given in Table I. The exponent 1/2 has been shown to result from coulomb gap at the fermi level\cite {efros75}. Parameters (i.e. \textit{m}, $T_o$) in PPy (powder) and PEDOT agree with those reported earlier\cite {amitava94,sanjib07}. Values of selected $\sigma_o(T)$'s are given in Table II. It is seen from Table I that of the first three samples, the sample 1 (PPy powder) has the least conductivity at room temperature while the sample 3 (PPy film) has the highest. We discuss the systems in this order as the maximum normalized conductivity (see below) also follows the same order.

\subsection{Polypyrrole (powder)}

The field dependence of dc conductivity $\sigma$ of a PPy(powder) sample at different temperatures ranging from 80K to 300K are shown in Fig. 2. The non-linear response of conductivity to the application of electric field can be clearly seen from the figure. The sample had a zero-bias linear conductance  $\Sigma_0$ of $0.003$ S at room temperature. A typical behavior at a constant temperature is that the conductivity remains constant for small fields and then starts increasing with increase in field. The value of the field at which conductivity starts deviating from its linear value $\sigma_o$ is the onset field $F_o$. A criterion to determine the latter is discussed below. With further increase in the bias, the conductivity continues to increase monotonically. As temperature is decreased, $\sigma_o$ decreases and the sample seems to become nonlinear at a field greater than the one required at a higher temperature i.e. $F_o$ \textit{increases} with decreasing temperature. This behavior is opposite to those found in other two systems (Figs. 4 and 6). Data at each temperature were first fitted to Eq. (\ref{eq:kaiser}) with all three parameters $\sigma_o$, $V_o$ and \textit{h} being free but fits were rather poor. In contrast, fits to the GM-expression (Eq. \ref{eq:gms}) containing terms up to \textit{n}=5 were reasonably good upto $\sigma/\sigma_o \sim 2.5$ and are shown in Fig. 2 (lines). But as seen particularly at \textit{T}=150 and 195 K, conductivity increased faster than the fitted curves at higher bias and could not be accounted even by including a \textit{n}=7 term. This may be related to incipient negative differential conductance regime as is evident in similar data in Fig. 2c of Ref. \onlinecite {ribo98}. Surprisingly, nonlinear least square fittings at all temperatures led to either very small or negative values for coefficients of even terms (i.e. \textit{n}=2,4) so that final fittings were done using only three terms: $\sigma(V) = \sigma_o + \sigma_3 V^{5/2} + \sigma_5 V^{14/3} $. Fitted values of the coefficients are given in Table II.

\begin{figure}
\includegraphics[width=6cm]{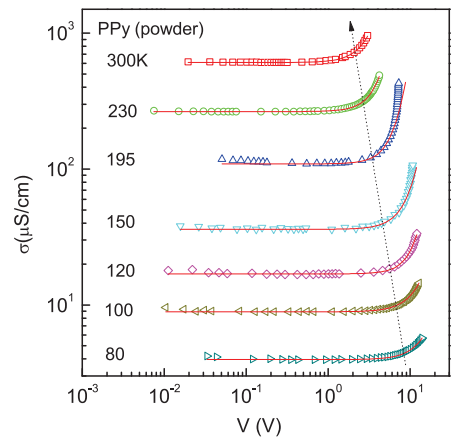}
\caption{(Color online) Variation of conductivity vs. electric field in a doped Polypyrrole pellet (sample 1) at different temperatures as indicated. The dotted line schematically indicates the movement of the onset bias with increasing temperature. The solid lines are fits to the Glazman-Matveev (GM) expression (Eq. \ref{eq:gms}). See text for details.}
\label{fig.2}
\end{figure}

\begin{figure}
\includegraphics[width=7cm]{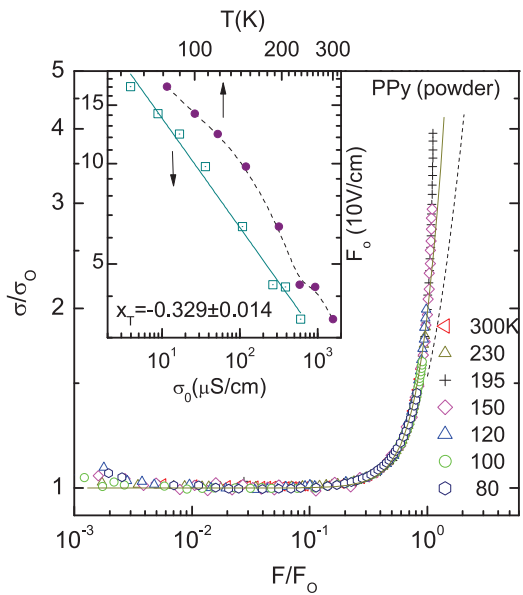}
\caption{(Color online) Scaling of the same data as in Fig. \ref{fig.2} to achieve data collapse as shown. The scale of the onset field $F_o$ is fixed by adopting a common criterion $\sigma(F_o)=2\sigma_o$ in this paper. The solid line is a fit to the scaled version (Eq. \ref{eq:scgms}) of GM expression with only two odd (\textit{n}=3,5) nonlinear terms: $\sigma/\sigma_o =1+ 0.54 q^{5/2}+ 0.42 q^{14/3}$ ($q=F/F_o$). The dashed line is a plot of the same expression without the \textit{n}=5 term. Inset shows two log-log plots of $F_o$ vs. $\sigma_o$ (open symbol) and \textit{T} (closed symbol). The solid line is a linear fit to the data with a slope $x_T$ as shown.}
\label{fig.3}
\end{figure}

Fig. 3 shows the result of making the data in Fig. 2 collapse into a single curve by suitable scaling. It is convenient to start with a temperature such that the data at that temperature are predominantly Ohmic but contain minimum non-Ohmic regime. In the present case the appropriate starting temperature is 80 K. The conductivity was scaled by its Ohmic values $\sigma_o$. For the field, any arbitrary choice (e.g. 1) for $F_o$ would do as far as data collapse is concerned. For the next higher temperature, the conductivity was scaled as before but $F_o$ was adjusted in such a way that this set of data merged with the earlier one as best as possible. The same procedure was then repeated for all the other temperatures in increasing order. Note that in this method $F_o$ is determined only upto a constant value. Multiplying all $F_o$'s by a constant only shifts the merged curve along the field axis without altering the curve anyway. To facilitate comparison, and for $F_o$ to be interpreted as an onset field, its scale was fixed by adopting an uniform criterion that the conductivity at the onset field would be double of its ohmic value i.e. $\sigma(F_o)= 2 \sigma(0)=2 \sigma_o$. The excellent data collapse up to about $\sigma/ \sigma_o \approx 3$ seen in Fig. 3 proves the existence of a field scale at each temperature. $F_o$ thus obtained following the above criterion is plotted with log-log axes as a function of both temperature $T$ (closed symbol) and the corresponding $\sigma_o$ (open symbol) in the inset. The solid line indicates a power law  $F_o \sim \sigma_o^{x_T}$ with an exponent $x_T$ being equal to $-0.329\pm 0.014$ which is negative as suggested by the orientation of the dotted line in Fig. 1. However, no such relation is apparent in the functional dependence on \textit{T}. Results thus validate the scaling as given in Eqs. (\ref {eq:scaling}) and (\ref {eq:fscale}). Since all the curves in Fig. 2, that were fitted by the GM-expression also collapse on to a single curve it is expected that the latter would also be fitted by the scaled form (Eq. \ref{eq:scgms}). This is indeed confirmed by the solid line in Fig. 3, which is a fit according to $\sigma/\sigma_o =1+ \overline{c}_3 {(F/F_o)}^ {5/2} + \overline{c}_5 {(F/F_o)}^{14/3}$ with $\overline{c}_3 = 0.54$ and $\overline{c}_5 = 0.42$. The relative high value of $\overline{c}_5$ confirms the rapid increase of the conductivity with field. A plot (dashed line) of the same fitting expression without the \textit{n}=5 term is also shown to highlight the contribution of the higher nonlinear term. The divergence of the data from the fitted curves at higher fields ($\sigma/ \sigma_o \approx 3$) as seen in the figure is due to th negative differential conductance as discussed above. $\overline{c}_n$'s in the fitting expression are simply averages of $c_n$'s calculated at each temperature from Eq. (\ref {eq:gmcoeff}) using $\sigma_n$'s and $V_o=0.1 F_o$ and given in Table II (for calculation of $\overline{c}_5$, data at 80 and 100 K were ignored as non-Ohmic regimes were small). The highest nonlinear term in the fitting expression depends upon the maximum value of the measured normalized conductivity, ${ (\sigma/\sigma_o)_{max} }$ which is about 4 in the present case. Notice that $\overline{c}_3 + \overline{c}_5 = 0.96 $ that is close to 1 as expected from Eq. (\ref{eq:cns}) by applying the criterion for $F_o$ which also leads to a condition for the scaling function in Eq. (\ref {eq:scaling}): \textit{g}(1)=2 seen also in Fig. 3. The deviation of the sum of $\overline{c}_n$'s from 1 is an indicator of how well the criterion for the field scale was implemented during scaling and quality of overall scaling.

It may be mentioned here that if one is interested only in the nonlinearity exponent and not the scaling function, it could be obtained  from experimental data by another method used by Gefen et al.\cite {gefen86} for characterizing the crossover to the nonlinear regime in a percolating system. In this method, one defines the crossover field $F_o$ such that the conductance at this field deviates by an arbitrarily chosen factor $\epsilon$ from its zero-field value i.e. $\Sigma(F_o) = \Sigma_o (1+\epsilon)$. Obviously, the value of the exponent should not depend upon the choice of $\epsilon$ as verified in discontinuous gold films\cite {gefen86}. Two values of $\epsilon$, 0.1 and 0.4, were considered but no significant variation in the exponent was observed. $\epsilon=1$ coincides with the criterion adopted in this paper. Clearly, this method will work as long as $\Sigma_o$ can be obtained from data. But particularly at low temperatures, Ohmic conductivities become too small to be above the noise floor of measurements. Hence, measurements are feasible only in highly nonlinear regimes and consequently, $\Sigma_o$'s can not be obtained directly from the data. In such cases, the method of scaling provides an alternative way to take into account such nonlinear data as illustrated in the cases discussed next.

\subsection{PEDOT (powder)}

The field dependence of dc conductivity $\sigma$ of a PEDOT(powder) system at different temperatures ranging from 4.9 to 26 K are shown in Fig. 4. The nonlinear response of conductivity to the application of electric field can be clearly seen from the figure. The sample had an Ohmic conductance $\Sigma_0$ of 0.026 S at room temperature. It is observed from the figure that the dotted line points towards right in contrast to the behavior in the previous case i.e. the onset field $F_o$ \textit {increases} with increasing temperature or conductivity. In fact, the orientation of the dotted line determines the overall shape of the curves which are somewhat divergent with bias in PPy (powder) (Fig. 2) but in contrast, appear convergent in PEDOT (powder) (Fig. 4). The converging feature naturally indicates eventual temperature independence of conductivity at large fields. As before, data at each temperature were fitted to the GM-expression (Eq. \ref{eq:gms}) containing terms up to \textit{n}=5. In this case, the least square method at all temperatures led to either very small or negetive values for coefficients of odd terms (i.e. \textit{n}=3,5) so that final fittings were done using only three terms: $\sigma(V) = \sigma_o + \sigma_2 V^{4/3} + \sigma_4 V^{18/5}$. The Ohmic conductivities at 4.9 and 6.5 K were obtained from the extrapolated line, $\ln \sigma_o$ vs. $T^{-0.5}$ in Fig. 1. The bias ranges at 10 and 12 K were too limited to yield reliable values for $c_4$'s. Fits (lines in Fig. 4) are seen to be excellent. Fitted values of the coefficients are given in Table II.

\begin{figure}
\includegraphics[width=6cm]{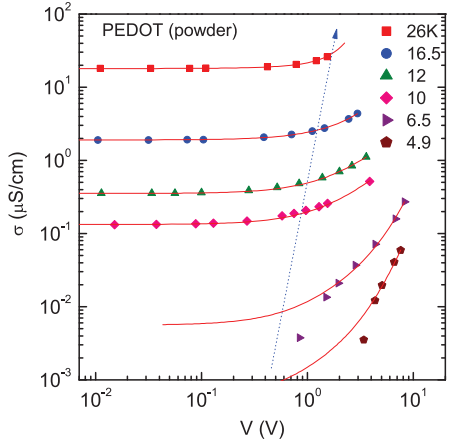}
\caption{(Color online) Variation of conductivity vs. electric field in doped PEDOT (sample 2) at different temperatures as indicated. The dotted line schematically indicates the movement of the onset bias with increasing temperature. The solid lines are fits to the GM-expression (Eq. \ref{eq:gms}). See text for details.}
\label{fig.4}
\end{figure}

\begin{figure}
\includegraphics[width=7cm]{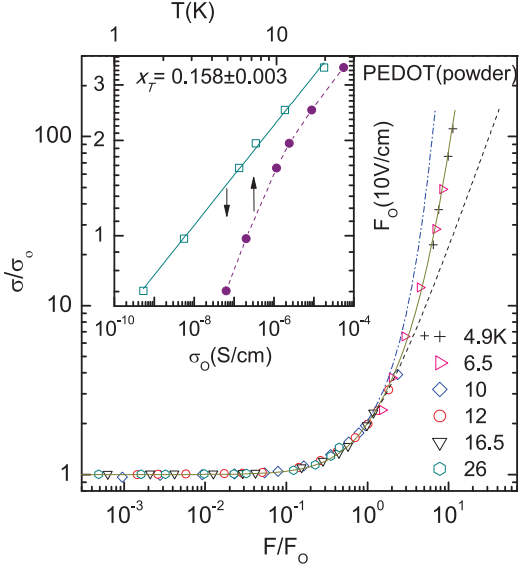}
\caption{(Color online) Scaling of the same data as in Fig. \ref{fig.4} to achieve data collapse as shown. The solid line is a fit to Eq.  (\ref{eq:scgms}) with only even (\textit{n}=2,4) nonlinear terms: $\sigma / \sigma_o =1+0.92 q^{4/3} +0.011 q^{18/5}$ ($q=F/F_o$). The dashed line is a plot of the same expression without the \textit{n}=4 term. The dash-dotted line is a fit to an exponential function (Eq. \ref{eq:lowF}). Inset shows two log-log plots of the scaling field, $F_o$ vs. $\sigma_o$ (open symbol) and \textit{T} (closed symbol). The solid line is a linear fit to the data (open symbols) with slope as shown.}
\label{fig.5}
\end{figure}

All the curves in the figure, starting with the one at 26 K and using the same criterion for $F_o$ as in the previous case, could be made to collapse into a single curve as shown in Fig. 5. Since the data for $T <7$ K have no linear region, it may be noted that the method in Ref. \onlinecite {gefen86} for determining the exponent can not be applied. However for scaling purpose,  extrapolated $\sigma_o$'s were used. However, when conductivities are scaled with these $\sigma_o$'s, the scaled data were all larger than the rest of data at higher temperatures. Absence of any overlapping data sets may leave some margin of error in $F_o$ which needs to be adjusted to complete the process of data collapse. The larger the gap between data sets is, the higher the margin of error is. The margin of error may be considerably reduced when, as in the present case, the collapsed curve is compared in somewhat self-consistent manner to some function it is expected to follow. The solid line in the figure is an excellent fit to $\sigma/\sigma_o =1+ \overline{c}_2 {(F/F_o)}^ {4/3} + \overline{c}_4 {(F/F_o)}^{18/5}$ with $\overline{c}_2 = 0.99$ and $\overline{c}_4 = 0.015$ so that $\overline{c}_2 + \overline{c}_4 =1.005$ as expected. $\overline{c}_2$ is the average of $c_2$'s (with data at \textit{T}=26 K ignored) whereas $\overline{c}_4$ is consistent with the values of limited number of $c_4$'s. A plot (dashed line) with only \textit{n}=2 term is also shown to highlight the contribution of the higher nonlinear term, which, as seen, is quite significant in this case although $\overline{c}_4 $ is quite small compared to $\overline{c}_2$. Note that ${ (\sigma / \sigma_o)_{max}}$ is about 100 compared to 4 in the previous case. Inset shows log-log plots of $F_o$ vs. temperature $T$ (closed symbol) and the corresponding $\sigma_o$ (open symbol). The solid line through open symbols indicates a power law with an exponent of $0.158\pm 0.003$. The exponent has a positive value in accordance with the orientation of the dotted line in Fig. 4. No reasonable straight line could be drawn through closed symbols.  

\subsection{Polypyrrole (film)}

Two sets of field-dependent conductivities of PPy films are presented in Fig. 6. The panel \textit{a} shows data taken in the sample 3 at different temperatures as marked and is similar to Figs. 2 and 4 whereas the panel \textit{b} shows data at \textit{T}$\approx$ 20 K taken in three different samples as indicated. In the former, the initial conductance changed due to change in the temperature whereas in the latter, the same was achieved by having different quenched disorder in the samples. As seen in the Figs. 2 and 6, the basic qualitative response to the electric field is similar in all the PPy samples irrespective of structure (i.e. powder or film) or sample condition in that at any temperature the conductivity behaves as a monotonically increasing function of field, starting from a constant value at small bias. However, a closer look reveals subtle differences as illustrated by the orientations of the schematic lines which indicate the movement of the onset field $F_o$ with increasing linear conductivity. In Fig. 6a, the line points towards right (i.e. $F_o$ \textit{increases} with the linear conductivity) as in Fig. 4 whereas in Fig. 6b, it points towards left (i.e. $F_o$ \textit{decreases} with the linear conductivity) as in powder (Fig. 2). Note that although both panels have one common set of data (sample 3 at T=20 K) this did not prevent two sets of data with two different driving variables - namely, temperature and disorder - from exhibiting opposite behavior in the onset field $F_o$. As mentioned in the previous case, the relation of the overall shape of the curves with the orientation of the dotted lines is now clearly seen in the two panels - the left one showing a convergent behavior as in Fig. 4 and the right one a divergent behavior as in Fig. 2 - in the \textit{same} system. At the low temperatures (2-5 K) conductivity indeed become nearly independent of temperature at large biases as seen in the panel \textit{a} - a feature that was strongly hinted in the data of PEDOT. The same is also apparent from roughly same values of $\sigma_6$ at low temperatures (see Table II). As before, data at each temperature or disorder were fitted to the GM-expression (Eq. \ref{eq:gms}) containing terms up to \textit{n}=6. In this case, the least square method at all temperatures led to either very small or negative values for coefficients of odd terms (i.e., \textit{n}=3,5) so that final fittings were done using only four terms: $\sigma(V) = \sigma_o + \sigma_2 V^{4/3} + \sigma_4 V^{18/5} + \sigma_6 V^{40/7}$. Fittings to data at $T \le 7$ K used the Ohmic conductivities obtained from extrapolation of $\ln \sigma_o$ vs. $T^{-0.5}$ line.

\begin{figure}
\includegraphics[width=8cm]{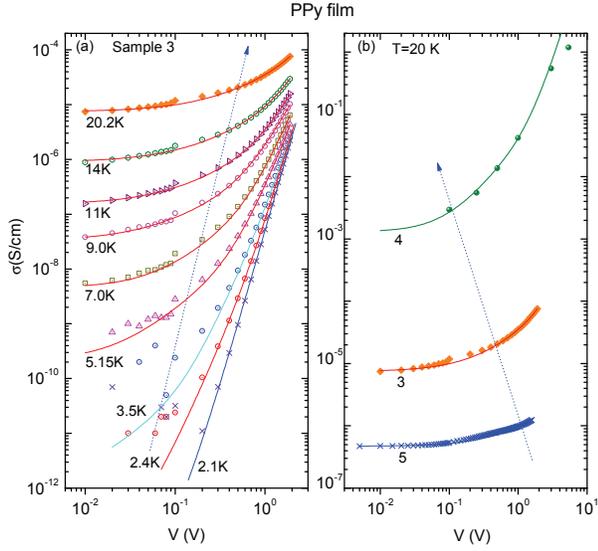}
\caption{(Color online) (a) Variation of conductivity vs. field in a doped PPy film (sample 3) at different temperatures as indicated. The dotted line schematically indicates the movement of the onset field with increasing temperature. The solid lines are fits to the GM expression (Eq. \ref{eq:gms}) with parameters given in Table II. (b) Similar data as on the left panel taken at the same temperature 20 K in three different samples as indicated with different quenched disorder. The dotted line schematically indicates the movement of the onset bias with increasing conductivity. The solid lines are fits using the GM model. See text for details. Note that data represented by solid diamonds in both panels are same.}
\label{fig.6}
\end{figure}

\begin{figure}
\includegraphics[width=8cm]{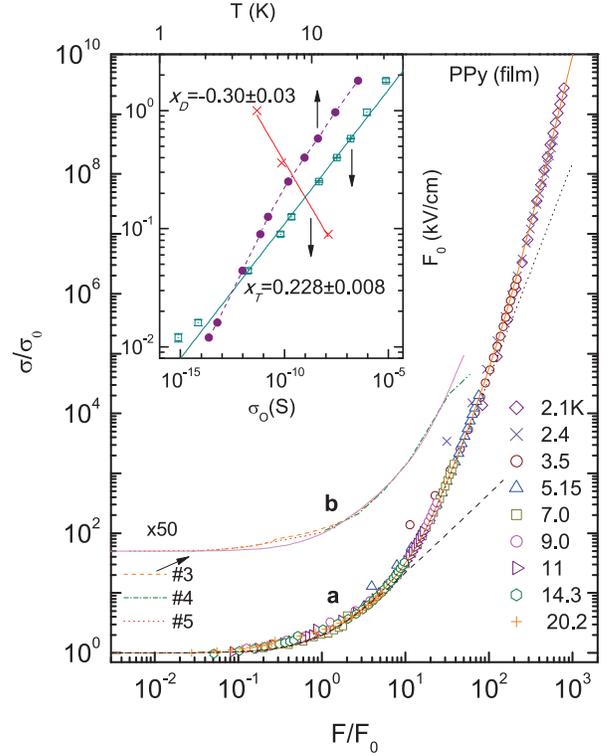}
\caption{(Color online) Scaled conductivity $\sigma/\sigma_0$ vs. scaled field $F/F_o$ of doped PPy films in two panels of Fig. 6. The scaled data \textbf{b} (lines) belonging to different samples are shifted upwards for clarity. The solid line is a fit to Eq. (\ref {eq:scgms}) with only two even (\textit{n}=2,4) nonlinear terms: $\sigma/\sigma_o =1+ 0.97 q^{4/3}+ 0.0013 q^{18/5} $ ($q=F/F_o$). The scaled data \textbf{a} with symbols belong to different temperatures. The solid line through it is a fit to the GM expression (Eq. \ref{eq:scgms}) with three even (\textit{n}=2,4,6) nonlinear terms: $\sigma/\sigma_o =1+ 0.97q^{4/3}+ 0.0024 q^{18/5}+ 6.5\time10^{-8} q^{40/7}$. The dashed line is a plot of the same expression but with terms up to \textit{n}=2 only. Similarly, the dotted line is a plot with terms up to \textit{n}=4 only. Inset shows three log-log plots of $F_o$ vs. $\sigma_o$ (open symbol and cross) and \textit{T} (closed symbol). The crossed symbols, belonging to three different samples, have been appropriately shifted (x$10^{-5}$, x0.2) to be within the scales as shown. The solid lines are linear fits to the data with slopes $x_T$ (open symbol) and $x_D$ (cross) as shown.}
\label{fig.7}
\end{figure}

Fig. 7 shows scaling of the data in two panels of Fig. 6. There are two curves showing data collapse - one (labeled \textbf{b}) belongs to different samples in panel \textit{b} and another (labeled \textbf{a}) belongs to the sample 3 at different temperatures in panel \textit{a}. Data collapse has been achieved by following the same procedure as adopted in previous cases including the criterion for fixing the onset field $F_o$, namely $\sigma(F_o)=2 \sigma_o$. Excellent data collapses are seen to have been achieved in both cases, with curve \textbf{a} covering nine orders of magnitude in conductivity and five orders in field. The data of curve \textbf{b} have been shifted upwards for clarity. Moreover lines, instead of symbols, are used in this case to highlight the quality of data collapse. In case of curve \textbf{a}, extrapolated $\sigma_o$'s were used for scaling particularly at low temperatures. The solid lines in the figure are excellent fits to $\sigma/\sigma_o =1+ \overline{c}_2 {(F/F_o)}^ {4/3} + \overline{c}_4 {(F/F_o)}^{18/5} + \overline{c}_6 {(F/F_o)}^{40/7}$. The coefficients $\overline{c}_2$, $\overline{c}_4$ and $\overline{c}_6$ are 0.98, 0.0024 and $6.5 \times10^{-8}$ for curve \textbf{a}, and 0.97, 0.0013 and 0 for curve \textbf{b}. In both cases, $\overline{c}_2 + \overline{c}_4 +\overline{c}_6 \approx 1$ as expected from Eq. (\ref{eq:cns}). $\overline{c}_2$'s are same as the average of $c_2$'s in Table II (with data at \textit{T}=2.1 K ignored) whereas $\overline{c}_4$'s are consistent with the values of limited number of available $c_4$'s. As with PEDOT, the values of $\overline{c}_2$'s are much greater than those of other coefficients although the latter's contributions are significant as indicated by the dashed and dotted lines. The latter are plots with up to \textit{n}=2 (dashed) and 4 (dotted) terms only. This is, of course, partly due to the criterion adopted for $F_o$. Note that ${ (\sigma / \sigma_o)_{max}}$ is about $10^9$ (curve \textbf{a}) compared to 100 and 4 in the previous cases. Inset shows log-log plots of $F_o$ thus obtained from scaling as function of both temperature $T$ (closed symbol, curve \textbf{a}) and the corresponding $\sigma_o$ (open symbol, curve \textbf{a} and cross, curve \textbf{b}). The solid line verifies the power law (Eq. \ref{eq:fscale}) with the exponent $x_T=0.228\pm 0.008$. No reasonable straight line could be drawn through the plot of $F_o$ vs. \textit{T} (closed symbols). In case of scaling with disorder, there are only three points that yield a tentative value of the exponent $x_D \sim -0.30$ (the subscript stands for disorder). As expected, $x_D$ compared to $x_T$, has a negative sign.

\begin{table*}
\caption{\label{tab:table2}%
Parameters in the GM-expression (Eq. \ref{eq:gms}) fitted to conductivity data at various temperature in five samples of three CP systems. $\sigma_n$ ($n \ge 2$) is in unit of S/cm ${\rm V}^{n-2/(n+1)}$. $c_n$'s are constants defined by Eq. (\ref{eq:gmcoeff}). A blank space indicates that the term was not used in the fit. A hyphen means that the data range did not permit reliable determination of the term. }
\begin{ruledtabular}
\begin{tabular}{cllccccccccc}

\textrm{Sample}&
\textrm{T(K)}&
\textrm{$\sigma_0$(S/cm)} &
\textrm{$\sigma_2$} &
\textrm{$\sigma_3$} &
\textrm{$\sigma_4$} &
\textrm{$\sigma_5$} &
\textrm{$\sigma_6$} &
\textrm{$c_2$} &
\textrm{$c_3$} &
\textrm{$c_4$} &
\textrm{$c_5(c_6)$} \\
\colrule
 
PPy     &80   &3.96$\times$10$^{-6}$  & & 2.10$\times$10$^{-9}$  & & 1.30$\times$10$^{-12}$ & & & 0.61  & & 0.19  \\ 
(powder)	&100  &8.96$\times$10$^{-6}$  & & 1.41$\times$10$^{-8}$  & & 6.50$\times$10$^{-12}$ & & & 0.66  & & 0.17  \\ 
			  &120  &1.69$\times$10$^{-5}$  & & 1.96$\times$10$^{-8}$  & & 8.62$\times$10$^{-11}$ & & & 0.40  & & 0.62  \\ 
 1		  &150  &3.67$\times$10$^{-5}$  & & 1.16$\times$10$^{-7}$  & & 2.88$\times$10$^{-10}$ & & & 0.58  & & 0.34  \\ 
        &195  &1.09$\times$10$^{-4}$  & & 6.90$\times$10$^{-7}$  & & 8.72$\times$10$^{-9}$  & & & 0.44  & & 0.49  \\ 
			  &230  &2.66$\times$10$^{-4}$  & & 3.23$\times$10$^{-6}$  & & 1.18$\times$10$^{-7}$  & & & 0.48  & & 0.42  \\ 
			  &300  &6.11$\times$10$^{-4}$  & & 1.63$\times$10$^{-5}$  & & 4.89$\times$10$^{-7}$  & & & 0.58  & & 0.25  \\ \cline{1-12}

PEDOT		&4.9  &5.30$\times$10$^{-10}$ &9.30$\times$10$^{-10}$ & &3.06$\times$10$^{-11}$ & & & 1.03  & & 0.014  \\
(powder)	&6.5  &5.60$\times$10$^{-9}$	&6.11$\times$10$^{-9}$  & &8.38$\times$10$^{-11}$ & & & 1.06  & & 0.014  \\
				&10   &1.33$\times$10$^{-7}$	&6.48$\times$10$^{-8}$  & &-        & & & 0.94  & & -      \\ 
 2			&12   &3.53$\times$10$^{-7}$	&1.38$\times$10$^{-7}$  & &-        & & & 0.96  & & -      \\ 
 				&16.5 &1.89$\times$10$^{-6}$	&5.23$\times$10$^{-7}$  & &4.53$\times$10$^{-9}$  & & & 0.94  & & 0.065  \\ 
 				&26   &1.80$\times$10$^{-5}$  &2.76$\times$10$^{-6}$  & &7.57$\times$10$^{-7}$  & & & 0.78  & & 3.447  \\ 	  \cline{1-12}

PPy 	  &2.1  &8.00$\times$10$^{-16}$ & -  & &$\sim$8.0$\times$10$^{-10}$ & &5.0$\times$10$^{-8}$ & -  & &$\sim$3.7$\times$10$^{-4}$ &(6.87$\times$10$^{-8}$)\\ 
(film)  &2.4  &7.00$\times$10$^{-15}$ &$\sim$1.8$\times$10$^{-11}$  & &2.3$\times$10$^{-8}$  & &5.8$\times$10$^{-8}$ &$\sim$1.13  & &0.0027 & (3.26$\times$10$^{-8}$)\\ 
      	&3.5  &1.75$\times$10$^{-12}$ &1.00$\times$10$^{-9}$ & &8.80$\times$10$^{-8}$ & &4.5$\times$10$^{-8}$ &1.05 & &0.0020  &(4.71$\times$10$^{-8}$)\\ 
      	&4.6  &6.34$\times$10$^{-11}$ &1.40$\times$10$^{-8}$ & &2.00$\times$10$^{-7}$ & &4.0$\times$10$^{-8}$ &1.06 & &0.0017  &(6.88$\times$10$^{-8}$)\\ 
      	&5.15 &2.20$\times$10$^{-10}$ &3.50$\times$10$^{-8}$ & &2.67$\times$10$^{-7}$ & &4.0$\times$10$^{-8}$ &1.19 & &0.0024  &(13.1$\times$10$^{-8}$)\\ 
 				&7    &4.53$\times$10$^{-9}$	&2.00$\times$10$^{-7}$ & &5.20$\times$10$^{-7}$ & &1.8$\times$10$^{-8}$ &0.83 & &0.0024 &(14.7$\times$10$^{-8}$)\\ 
 3			&9    &3.29$\times$10$^{-8}$	&9.19$\times$10$^{-7}$ & &7.84$\times$10$^{-7}$ & &3.0$\times$10$^{-9}$ &0.56 & &0.0045 &(4.96$\times$10$^{-8}$)\\ 
 				&11   &1.50$\times$10$^{-7}$	&1.54$\times$10$^{-6}$ & &1.84$\times$10$^{-6}$ &  &- &0.56 &  &0.0053 &- \\ 
 				&14.3 &9.00$\times$10$^{-7}$  &5.83$\times$10$^{-6}$ & &2.59$\times$10$^{-6}$ &  &- &0.73 &  &0.0079 &- \\
 				&20.2 &7.47$\times$10$^{-6}$  &2.48$\times$10$^{-5}$ & &3.09$\times$10$^{-7}$ &  &- &0.86 &  &0.0105 &- \\ \cline{1-12}	 

 4			&20.2 &1.30$\times$10$^{-3}$  &3.01$\times$10$^{-2}$ & &1.03$\times$10$^{-2}$ &  &- &0.94 & &0.0014 &- \\ \cline{1-12}
 5			&20   &4.70$\times$10$^{-7}$  &5.23$\times$10$^{-7}$ & & - & &- &1.10 & &- &- \\
 				
\end{tabular}
\end{ruledtabular}
\end{table*}

\subsection{Other systems in literatures}

Nonlinearity exponents $x_T$ in the three systems discussed above are shown in Table III. The latter also contains, for sake of comparison, results from digitized data of three other CP systems available in literatures. All these systems invariably exhibit the property of scaling (Eq. \ref {eq:scaling}). The systems include a \textit{p}-Toluensulfonate (PTS)-doped PPy (PPy(R)) film\cite {ribo98}, PTS-doped polydiacetylene (PDA) single crystal\cite {aleshin04} and iodine-doped polyacetylene (PA) nanofibre\cite{kaiser04a}. The PPy(R) film (0.1-0.15 mm thickness) was obtained by electro-deposition at the current density of 0.2 mA/cm$^{-2}$. The scaled curve of conductivities at four temperatures (16-31 K) in this film looked very similar to that in Fig. 3. It also yielded a negative exponent, -0.16 compared to -0.33 obtained in PPy (powder). The temperature-variation of the Ohmic conductivity of this system has been mentioned in the introduction.

\begin{figure}
\includegraphics[width=6cm]{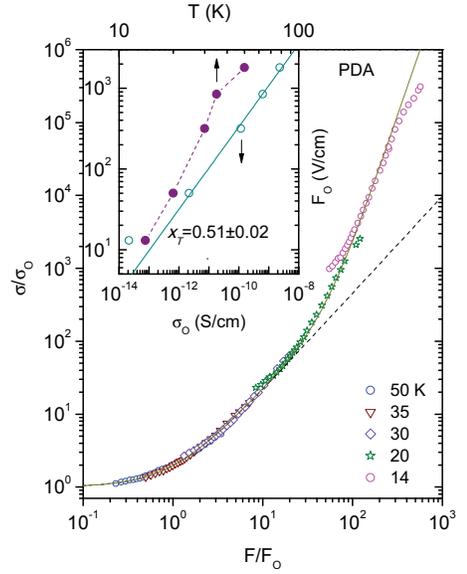}
\caption{(Color online) Scaling of the data in a PDA single crystal from Ref. \onlinecite {aleshin04} to achieve data collapse as shown. The solid line is a fit to Eq. (\ref{eq:scgms}) with only even (\textit{n}=2,4) nonlinear terms: $\sigma / \sigma_o =1+0.98 q^{4/3} +0.000123 q^{18/5}$ ($q=F/F_o$). The dashed line is a plot of the same expression without the \textit{n}=4 term. Inset shows two log-log plots of the scaling field, $F_o$ vs. $\sigma_o$ (open symbol) and \textit{T} (closed symbol). The solid line is a linear fit to the data (open symbols) with slope as shown.}
\label{fig.8}
\end{figure}

PTS-doped PDA crystals are quasi-1D in nature, consisting of weakly coupled linear parallel chains of covalently bonded carbon atoms. It follow a VRH-type conduction (\textit{m}=0.65-0.70, $T_o=2570$ K) at low temperatures with a crossover at about 50 K to activated conduction at higher temperatures. Five temperatures between 50 and 14 K were used such that corresponding data could be digitized with some reasonable accuracy from linear current-field plots (viz. Fig. 7 of [\onlinecite {aleshin04}]) which, particularly at low bias, are more prone to digitizing errors than conductivity-field ones. Nonetheless, Fig. 8 shows clear evidence of scaling in conductivities at different temperatures covering nearly six decades in conductivity and four decades in field. The solid line in the figure is an excellent fit over the whole range to $\sigma/\sigma_o =1+ \overline{c}_2 {(F/F_o)}^ {4/3} + \overline{c}_4 {(F/F_o)}^{18/5}$. The coefficients $\overline{c}_2$ and $\overline{c}_4$ were 0.98 and 0.000123 respectively and add up to nearly 1 as expected. These values were consistent with the coefficients obtained from fittings of data at each temperature to the GM-expression (Eq. \ref{eq:gms}). The dashed curve is a plot with up to \textit{n}=2 term only. It was found that fitted values of $\sigma_o$'s (also used in scaling) were progressively less than those given in the paper with decreasing temperature, by as much as an order of magnitude at 20 K. ${ (\sigma / \sigma_o)_{max}}$ is about $10^5$ compared to $10^9$ in PPy film (this work). The nonlinearity exponent was $x_T=0.51 \pm 0.02$. It is seen in Fig. 8 that curves particularly at lower temperatures tend to rise less rapidly at higher fields than lower ones. This is because those portions of the curves follow $F^{-1/2}$ dependence (Eq. \ref {eq:highF}) and are outside of the scaling domain\cite {taluk10}. In fact, the collapsed curve really represents an envelope of all scaled curves at different temperatures. Such $F^{-1/2}$ dependence in PPy film of this work was insignificant (see Fig. 7) although both these systems had the VRH exponent \textit{m} equal to 1/2.

\begin{figure}
\includegraphics[width=6cm]{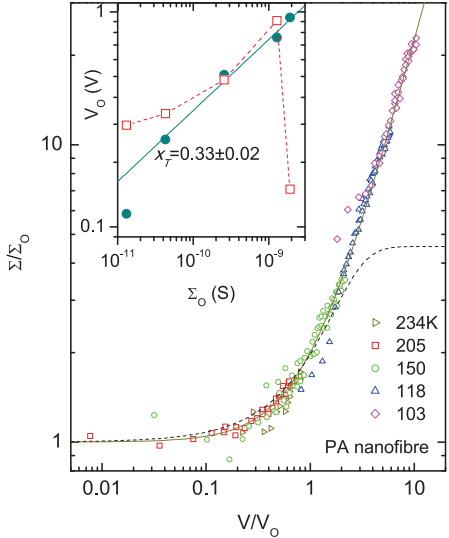}
\caption{(Color online) Scaling of the data in a PA nanofibre from Ref. \onlinecite {kaiser04a} to achieve data collapse as shown. The solid line is a fit to the GM-expression (Eq. \ref{eq:scgms}): $\sigma / \sigma_o =1+ (V/V_O)^{4/3}$ . The dashed line is a plot of the expression (Eq. \ref{eq:kaiser}) with \textit{h}=0.21. Inset shows two log-log plots of the scaling bias $V_o$ vs. $\Sigma_o$. One set of data (open symbols) is obtained from fittings to Eq. (\ref{eq:kaiser}) and another set (closed symbol) from scaling. The solid line is a linear fit to the data (closed symbols) with slope as shown.}
\label{fig.9}
\end{figure}

Individual iodine-doped PA nanofibres of diameters 10-40 nm were investigated for transport properties. At low temperatures below 30 K, the \textit{I-V} characteristics follow the Zener-type tunneling, $ \Sigma = \Sigma_o \exp (-F_u/F) $ where $F_u$ in the exponent depends on the magnitude of the energy gap and the effective mass of tunneling electrons\cite {jgpark2}. The \textit{I-V} curves were non-Ohmic and temperature independent up to 30K above which \textit{I-V} curves were temperature dependent. Upon comparing with Eq. (\ref{eq:highF}) it becomes evident that the above equation really characterizes the high-field regime that becomes apparent particularly at low temperatures (see also Fig. 6) and is not compatible with the scaling (Eq. \ref{eq:scaling}). At higher temperatures the Ohmic conductance exhibits an activated-type conduction and \textit{I-V} curves again exhibit the same scaling behavior shown in Fig. 9 as in other cases discussed here. \textit{I-V} data were obtained in a single iodine-doped PA nanofibre of diameter ~20 nm at various temperatures (Fig. 1 of [\onlinecite {kaiser04a}]). Out of those, data at five temperatures (234-103 K) were found suitable for digitization and presented in scaled form in Fig. 9. In spite of some inherent noise in the data, high quality of data collapse is quite evident. The solid line in the figure represents an excellent fit to $\sigma/\sigma_o =1+ {(F/F_o)}^ {4/3}$ thus requiring only \textit{n}=2 term in the GM-expression. Authors\cite {kaiser04a} fitted the \textit{I-V} curves to the expression (Eq. \ref{eq:kaiser}). The dashed curve  is a plot of the Kaiser expression (Eq. \ref{eq:kaiser}) with \textit{h}=0.21. The particular value of \textit{h} was chosen to satisfy the criterion for the bias scale, $\Sigma (V_o) = 2 \Sigma_o$. The fit runs above the data for $V/V_o < 1$, and runs below the data for $V/V_o > 1$. Moreover, the variation in \textit{h} with temperature and the saturation at large bias clearly reveals inadequacies of the Kaiser expression (Eq. \ref{eq:kaiser}) in describing the field-dependent conductance. Discrepancies at low bias were already noted\cite {zhi09}. Inset shows two plots of $V_o$ vs. $\Sigma_o$ with one set of $V_o$'s (closed symbols) obtained from scaling and another set of $V_o$'s (open symbols) obtained from fittings to Eq. (\ref {eq:kaiser}). Comparison of the two sets shows further limitations of the Kaiser expression. The nonlinearity exponent was determined to be $x_T=0.33 \pm 0.02$.

\section{Discussions}

The last section dealt with four different CP systems, namely PPy, PEDOT, PDA and PA in as many forms, namely powder, film, crystal and nanofibre. Even in such diverse conditions, the scaling phenomena as embodied in Eq. (\ref {eq:scaling}) and demonstrated in Figs. 3,5,7,8 and 9 stands validated in clear and unambiguous manner. Furthermore, the figures also confirm the remarkable fact that there exists a \textit{single} field scale in any given sample at least within the experimental ranges of field and conductivity spanning more than five and nine orders of magnitude respectively. This is contrary to the predictions in the field-dependent VRH theories discussed in section II. Thus, an important objective of this paper, as stated in the introductory section, is fulfilled. We believe that scaling phenomena observed in CP's here, and in amorphous- and doped-semiconductors\cite {taluk10} earlier, thus indicate a general and fundamental property of the class of disordered systems with localized states. The scaling analysis here follows that of critical phenomena in thermodynamic phase transitions\cite {stanley} and hence, the method of analysis described in section III naturally differs from that hitherto adopted e.g. in Refs. \onlinecite {bufon07}, \onlinecite {aleshin04} and \onlinecite {ladieu00}. The method, in absence of a proper theory, is primarily phenomenological but finally yields a concrete number - the nonlinear exponent - as a characterization of the underlying conduction mechanism. Let us now consider details of scaling, namely the scaling variables (which $F_o$ depends on) and scaling function, \textit{g} and their possible connections to the microscopic picture.

\subsection{Field scale $F_o$ and scaling variable}

To start with, let us note that at low fields, the variable range hopping as the conduction mechanism is often observed in conducting polymer systems (barring systems like PA nanofibre). VRH is actually a phonon-activated hopping between localized states irrespective of the presence or absence of polarons. In such situations, the conductivity as a function of temperature is generally described by Eq. (\ref{eq:mott}) with \textit{m} ranging from 1/4 to 1. Discussed in section II, traditional theories incorporating field effects in VRH conduction have a number of predictions or implications which are at variance with experimental results:

First, two field scales $F_o$ and $F_u$ (corresponding to the two length scales, the hopping length $R_h$ and localization radius \textit{a} respectively) are predicted whereas only one scale is experimentally observed; Second, $F_o$ is basically set by the temperature scale and supposed to vary as $F_o \sim T^{\alpha}$ (Eq. \ref {eq:FT}) where $\alpha = 1+m\mu$ is a positive number and always greater than 1. But log-log plots of $F_o$ vs. \textit{T} (insets in Figs. 3,5,7 and 8) generally deviate from linearity in varying degrees. They seem to be better described by two power-laws with two exponents, $\alpha_1$ at low temperatures and $\alpha_2$ at higher temperatures with $\alpha_1 \ge \alpha_2$. Nonetheless, if the data are still subjected to linear fittings, slopes (i.e. $\alpha$'s) turned out to be 0.98, 2.24 and 4.1\cite {paper25_1} in PEDOT, PPy (film) and PDA respectively compared to predicted values of 1.5(2), 1.5(2) and 1.68(2.36) for $\mu$=1(2) respectively; Third, the problem of temperature directly determining the field scale is rather dramatically highlighted by the measurements in three samples of PPy film (Fig. 6b) at a temperature of 20 K, each one having a different quenched microscopic disorder and characterized by the Ohmic conductivity $\sigma_o$. If $F_o$ is indeed set by temperature alone, it should be basically same for each sample in the figure as measurements were performed at the same temperature. However, Inset in Fig. 7 clearly shows that $F_o$ varies with $\sigma_o$ and within the limited range, is compatible with Eq. (\ref{eq:fscale}) albeit with a different nonlinearity exponent. It proves that a field scale is an intrinsic property of a disordered sample and is a function of various parameters like disorder, temperature etc.. Significance of the role of disorder becomes further apparent when compared with the results obtained by Bufon et al.\cite{bufon07} from measurements at different magnetic fields. Fig. 11 displays \textit{$\sigma$-V} curves (dashed) at three magnetic fields \textit{B} as indicated. The conductivity generally decreases with increasing magnetic field. It is found that the three curves could be made to collapse into a single curve (solid) by simply scaling the conductivity at each magnetic field by a factor $\lambda$ (arbitrary up to a constant factor) as shown. The bias did not need any scaling. This means that the field scale was independent of \textit{B} unlike disorder, and consequently, the corresponding nonlinearity exponent $x_B$ is 0; Fourth, in some systems like PPy (powder) (inset of Fig. 3) and PPy(R) film (Table III), $\alpha$ (as well as the nonlinearity exponent $x_T$) is negative. This is a serious discrepancy as it is irreconcilable with the theories.

\begin{figure}
\includegraphics[height=6.5cm]{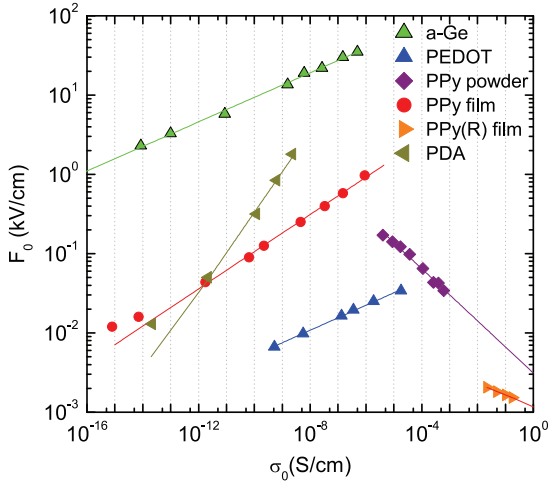}
\caption{(Color online) Field scale vs. linear conductivity in various systems. The plot for \textit{a}-Ge is from Ref. \onlinecite{taluk10}. The data of PPy(R) film have been shifted to the left by a factor of $10^4$ to lie within the scales. }
\label{fig.10}
\end{figure}

In contrast, log-log plots of $F_o$ vs. $\sigma_o$ are consistently linear as seen in Figs. 9 and 10, the latter displaying data from different CP systems (Figs. 3,5,7 and 8), including PPy(R) in the same scales (PA nanofibre is excluded because of variables with different units). Linearities in the plots give credence to the empirical power-law $F_o \sim {\sigma_o}^{x_T}$ (Eq. \ref{eq:fscale}) which is radically different from Eq. (\ref {eq:lowF}) in that the variable in the power-law is not the temperature itself but the linear conductivity $\sigma_o$ which is, of course, a function of temperature. All these considerations suggest that $\sigma_o$ should be considered as the appropriate scaling variable in Eq. (\ref{eq:scaling}) which then reads as
\begin{equation}
{\sigma(M,F) / \sigma_o}  = g (F {\sigma_o}^{-x_M} ),
\label{eq:scaling2}
\end{equation}
where \textit{M} stands for the variable(s) used to vary $\sigma_o= \sigma(M,F=0)$. Eq. (\ref{eq:scaling2}) allows description of the scaling of the field-dependent conduction along various paths in the variable space in a natural fashion without any obvious contradiction. In analogy with the scaling formulation of thermodynamic critical phenomena\cite {stanley}, $\sigma_o= 0$ defines a 'critical' point. The Ohmic conductivity $\sigma_o$ plays the traditional role of temperature \textit{T} in that it sets the field scale which, in turn, must be given by some physical length scale. The latter is yet to be explained but must be distinct from either \textit{a} or $R_h$. Similar problem of unsolved length scale\cite {lee88} exists also on the metallic side of the metal-insulator transition \cite{yoon94} in doped semiconductors (e.g. Si:P). The linear conductivity of a sample at 0 K on the metallic side goes to zero at the transition as a power-law as the doping level is decreased to a critical value. Such critical behavior is associated with a diverging length scale that still remains unsolved. However, one must be careful not to carry the analogy with temperature too far as the linear conductivity is a non-equilibrium quantity which may behave differently than the equilibrium quantities\cite {mb03}. It may be noted that the right hand side of Eq. (\ref{eq:fscale}) could still contain a slow varying function of temperature or some other variable but the dominant variation would still be given by the power-law. The doping level like the conducting fraction \textit{p} in composites is the natural quantity for parameterization of quenched disorder in CP's. However, in practice the doping level often may not be known. Under this situation, we use resistivity $\rho_o(T) = 1/ \sigma_o(T)$ as a measure of disorder. This is intuitively satisfying but has the drawback of being temperature-dependent.

\begin{figure}
\includegraphics[width=6cm]{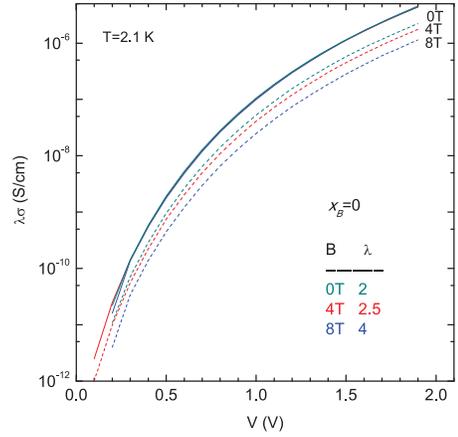}
\caption{(Color online) Conductivity vs. bias voltage (dashed curves) in the PPy film (sample 3) at three magnetic fields (after Ref. \onlinecite {bufon07}). Only conductivity of each curve need to be scaled with a factor $\lambda$ to collapse on to the solid curve. The non-linearity exponent $x_B$ is zero.}
\label{fig.11}
\end{figure}

While allowing a view of data ranges in six samples at a glance,  Fig. 10 also provides a basis for quantitative comparison of the field scales across those systems as well as amorphous germanium (\textit{a}-Ge)\cite {taluk10} included for reference. The comparison, of course, is not straightforward as the slopes $x_T$ have different values. However, it is seen that CP systems are generally electrically 'soft' compared to \textit{a}-Ge since the onset field in a CP system is smaller than that in \textit{a}-Ge in the displayed range of conductivity. For a quantitative comparison, let us consider \textit{a}-Ge and PEDOT since both systems have nearly same slope (nonlinearity exponent) of 0.16. The prefactor $A_T$ in Eq. (\ref{eq:fscale}) which represents the strength of a field scale is 331 and 0.2 (kV/cm)(S/cm)$^{-0.16}$ respectively i.e., the onset field in a-Ge is roughly 1600 times that in PEDOT. Now, simple dimensional considerations lead to an expanded expression for $F_o$:
\begin{equation}
F_o = C_1 { k_B T_o \over ea} {\left ( {ah \sigma_o \over e^2} \right )}^{x_T},
\label{eq:ffscale}
\end{equation}
where $C_1$ is a dimensionless unknown constant, $e^2/h$ is the quantum of conductance. Accordingly, $A_T \sim T_o /a$. Taking localization lengths in the two systems to be of same order of magnitudes ($\sim$1 nm) and with $T_o \approx 10^8$ and 1800 in \textit{a}-Ge and PEDOT respectively, the ratio $F_o$(\textit{a}-Ge)/$F_o$(PEDOT) is about $\sim 10^5$, two orders of magnitude greater than 1600. VRH theories fare even worse. According to Eq. (\ref {eq:FT}), $F_o \sim 1/a{T_o}^{\alpha -1}$ so that the ratio $F_o$(\textit{a}-Ge)/$F_o$(PEDOT) is about $1800^{1/2}/ 10^{8/4} \sim 1$, three orders of magnitude less than the experimental value.

\subsection{Scaling function}
At least three forms of nonlinear \textit{I-V} curves in hopping systems have been discussed in section II. The field-dependent VRH theories\cite{hill71,pollak76,shklov76} predict a simple exponential (Eq. \ref{eq:lowF}) at low and intermediate fields, to which the scaled data of PEDOT(powder) have been fitted for illustration purpose (dash-dotted line in Fig. 5). The fitted curve appears to match data well at least numerically in the low field region ($F/F_o \sim 1$) but gradually deviates (rising faster than data) from it in the high field region as predicted. Nevertheless, as shown in the inset of Fig. 5, $F_o$ fails to follow the temperature-dependence as given by Eq. (\ref {eq:FT}). Such inconsistent behavior of the VRH theory is quite typical. Inadequacies of the Kaiser expression (Eq. \ref{eq:kaiser}) in case of PA nanofiber have been discussed earlier in details (see Fig. 9). Interestingly, fits\cite {kaiser04a} to Eq. (\ref{eq:kaiser}) were visually quite acceptable. This brings us to the GM-expression (Eq. \ref{eq:gms}) which is used for the first time to describe the CP data in various systems. 

Fits to the \textit{I-V} data at different temperatures as well as disorder are shown in Figs. 2,4 and 6, and to the scaled curves in Figs. 3,5,7,8 and 9. The near perfect agreement between the experimental data and theoretical fits is quite remarkable in view of range of data covered - more than nine orders of magnitude in conductance and nearly five orders of magnitude in field. Furthermore, fits to the scaled curves using the scaled version of the GM-expression (Eq. \ref{eq:scgms}) provide absolute justification of the assumption (Eq. \ref{eq:gmcoeff}) and underline the relations that exist among the coefficients in the GM expression but were not foreseen in the theory\cite {matveev88}. The fact that coefficients $\overline{c}_n$'s used in fits to the scaled curves correspond closely (within errors) to $c_n$'s (Table II) obtained from fittings of individual curves also demonstrate those relations in a self-consistent manner. It may be recalled that two assumptions about the coefficients, namely Eqs. (\ref{eq:gmcoeff}) and (\ref{eq:yn}) led to the important expression Eq. (\ref{eq:ne2}) for the non-linearity exponents $x_T$ in the GM-model, that allowed a number of conclusions to be drawn on the properties of $x_T$ itself. Fitted values of the coefficients $\sigma_n$'s (Eq. \ref{eq:gms}) for different values of \textit{n} given in Table II are plotted against $\sigma_o$ using log-log scales in Fig. 12. The linearity of the plots amply validate the assumption (Eq. \ref{eq:yn}). Interestingly, the latter, under certain condition, can be derived within the original context of GM-model\cite {matveev88}. According to the theory, coefficients $\Sigma_n \sim \exp[-2w/a(n+1)]$ are exponentially increasing albeit slowly varying functions of \textit{n} where \textit{w} is the typical barrier thickness. The Ohmic conductance $\Sigma_o$ is sum of two terms - $\Sigma_d \sim \exp (-2w/a)$ from direct tunneling and $\Sigma_1 \sim \exp (-w/a)$ from resonant tunneling. For thick barrier, $\Sigma_o \sim  \exp (-w/a)$ so that $\Sigma_n \sim {\Sigma_o}^{y_n}$ with 
\begin{equation}
y_n = {2 \over n+1},   \label{eq:ynexp}
\end{equation}
for $n \ge 2$. Thus, $y_n = 0.67,~0.5,~0.4$ ... does satisfy inequalities for $x_T>0$ (Eq. \ref{eq:y1yn}). In fact, the value of $y_2$ i.e. 0.67 is close to some experimental values in Table III but seems to be much slowly decreasing than required. More importantly, there is a spectrum of values rather than a single value for a given channel.

\begin{figure}
\includegraphics[width=7cm]{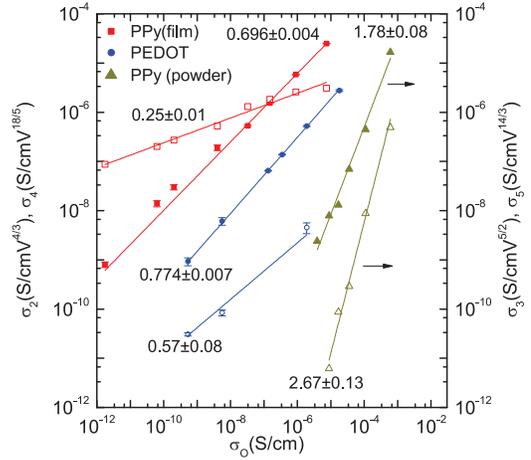}
\caption{(Color online) Log-log plots of GM coefficients vs. linear conductances. Solid lines are linear fits to the data with slopes as indicated.}
\label{fig.12}
\end{figure}

A close look at Table II reveals yet another intriguing feature of channel selection - only the 'even' channels (i.e., channels with even number of localized states) appear in the fittings to systems  like PEDOT, PPy film etc. having positive nonlinearity exponents ($x_T > 0$) whereas only the 'odd' channels appear in the fittings to other systems like PPY (powder) having negative non-linearity exponent ($x_T<0$). As a result, the lowest channel contributing to non-Ohmic conduction in the former is a two-impurity channel ($n_o=2$), and a three-impurity channel ($n_o=3$) in the latter. There was no instance of mixing of terms of the two series (i.e. even and odd) in any of the samples investigated here. Interestingly, no such selection of channels was detected in a model system of metal-amorphous silicon-metal tunnel junctions where all channels from \textit{n}=2 to 5 were found to be present in some \textit{$\Sigma$-V} curves\cite {beasley95}. The theory, at present, does not provide any clue to this phenomena of selection, in general, and to the properties of CP systems responsible for this, in particular. Similar selection has been also noticed in other hopping systems like carbon nanofibers and amorphous/doped semiconductors\cite {taluk10}. We also note an opposite trend for PPy films with different quenched disorder (fig. 6b) having a negative nonlinearity exponent where the best fits are obtained with even channels in the GM expression. This is not surprising as one of the curves was earlier part of the data collapse of field dependent conductances at various temperatures. The non-linearity exponent was found positive (inset, fig. 6a) and fits to the GM-expression consisted of even channels only. Differences between the systems described by even and odd channels can be seen in the values of $c_n$'s in Table II. In case of even channels, $c_2$'s are close to 1 and rest of $c_n$'s are very small. But in case of odd channels, $c_n$'s are comparable as in PPy (powder). 

It is seen from Figs. 5,7 and 8 that the CP's like PEDOT, PPy and PDA were described by the GM expression with only even non-Ohmic channels: $\sigma / \sigma_o = 1+ \overline{c}_2 q^{4/3} + \overline{c}_4 q^{18/5} + \overline{c}_6 q^{40/7} $ where $q=F/F_o$. The data ranges were sufficiently large that each system needed at least two non-Ohmic terms for fitting. While the coefficient $\overline{c}_2$ was close to 1 in all the systems, $\overline{c}_4$ was 0.011, 0.0024 and 0.000123 respectively. Such decreasing values of $\overline{c}_4$ means decreasing rate of rise of conductivity with field. One indicator of this trend could be $q_4$, the value of \textit{q} in a system above which the fitting (the dashed curve in the corresponding figure) with only the lowest channel (\textit{n}=2) deviates from the scaling curve. The increasing value of $q_4$, roughly 2,10 and 20 respectively in the three systems do correlate with $\overline{c}_4$. More significantly, the decreasing $\overline{c}_4$ correlates well with the increasing nonlinearity exponent which are 0.16, 0.23 and 0.51 respectively (Table III): smaller the value of a nonlinearity exponent is, the steeper is the conductivity curve (or larger the value of $\overline{c}_4$). This correlation, in fact, is a result of the relation Eq. (\ref {eq:large}). The higher order coefficients like $\overline{c}_6$ are expected to exhibit similar correlation in values. If one adopts the view that each distinct value of the nonlinearity exponent constitutes a different universality class then one appreciates the unique role that the values of $\overline{c}_n$'s play in giving identity to the corresponding scaling function. Incidentally, $q_4$ is given by $\sigma / \sigma_o = 1 + \overline{c}_2 {q_4}^{p_2} $ from Eq. (\ref{eq:scgms}). $q_4$ may be interpreted to be the onset field for the 4th channel.
 
It is quite remarkable that only three nonlinear terms were required to describe data in PPy (film) over nine decades. The finite number of terms is consistent with the requirement that the GM series must terminate at some finite \textit{n} (the point iv in section II). The highest term with \textit{n}=6 ($p_n=5.7$) was necessary to fit the data at high field ($ F/ F_o \gg 1$) i.e. data at $T \sim 2.1$ K which is nearly given by a power-law (Fig. 6a). Actually a log-log plot of the data yielded a slope of $z_T=5.57$ close to 5.7. However this value of $z_T$ is higher than $1/x_T = 1/0.23 \approx 4.39$ in apparent disagreement with Eq. (\ref {eq:large}) for $x_T>0$. The same happens also in PDA where $z_T (\sim 3.6) > 1/x_T =2$. Reasons for this quantitative discrepancies are not clear although qualitative compliance is obvious as discussed above. The relation (Eq. \ref{eq:large}) has been routinely observed to hold in amorphous semiconductors \cite{taluk10}. It is to be noted that in both PPy (film) and PDA, $F_o$ at low temperatures lie above the linear fits (Fig. 10). Whether this is indicative of existence of two slopes or not merits further careful measurements at low temperatures to resolve this issue.

\begin{table*}
\caption{\label{tab:table3}%
Comparison of nonlinearity exponents $x_M$ obtained using scaling and extended GM analysis (Eq. \ref{eq:ne2}) in various CP systems. \textit{M} stands for the variable used to vary $\sigma_o$: \textit{T}-Temperature, \textit{D}-Disorder and \textit{B}-Magnetic field. $y_n$ is given by $\sigma_n \sim {\sigma_0}^{y_n}$ with values of $\sigma_n$'s of some samples displayed in Table II.}
\begin{ruledtabular}
\begin{tabular}{lccccccccr}

\textrm{System}&
\textit{M}&
\textrm{$x_M$}&
\textrm{$x_M$}&
\textrm{$y_2$}&
\textrm{$y_3$}&
\textrm{$x_M$}&
\textrm{$y_4$}&
\textrm{$y_5$}&
\textrm{$y_6$} \\
   &
   &
\textrm{scaling} &
\textrm{GM(\textit{n}=2/3)}   &
   &
   &   
\textrm{GM(\textit{n}=4/5)}   &
   &
   & \\
\colrule
PPy (powder)   &\textit{T} &$-0.329\pm0.014$ &$-0.31\pm0.03$ & &$1.78\pm 0.08$ & $-0.36\pm0.03$ & &$2.67\pm 0.13$ \\
PPy (film)\cite {ribo98} &\textit{T}  &$-0.155\pm0.012$ &$-0.13\pm0.02$ & & $1.32\pm 0.04$  \\  
PPy (film)   &\textit{T}  &$0.228\pm 0.008$ &$0.230\pm 0.003$ &$0.693\pm 0.004$& &$0.21 \pm 0.01$ &$0.25 \pm0.01$& & $\sim$0 \\ 
PPy (film)   &\textit{D}  &$\sim -0.3$ &  &   & &  &  & &  \\ 
PPy (film)   &\textit{B}  & 0 & 0 & 1& & 0 &1 & & 1 \\ 
\cline{1-10} 
PEDOT	(powder)		&\textit{T}  &$0.158 \pm 0.003$ &$0.170 \pm 0.005$ & $0.774\pm 0.007$& &$\sim 0.12 $ & $\sim 0.57 $   \\  \cline{1-10}  
PDA (crystal)\cite{aleshin04} &\textit{T} &$0.51 \pm 0.02$ &$0.49\pm0.02$ &$0.34 \pm 0.02$  \\  \cline{1-10}  
PA (nanofibre)\cite{kaiser04a} &\textit{T} &$0.33 \pm 0.02$  &$0.29 \pm 0.06$ &$0.62 \pm 0.08$    \\ 
\end{tabular}
\end{ruledtabular}
\end{table*}

\begin{figure}
\includegraphics[width=6cm]{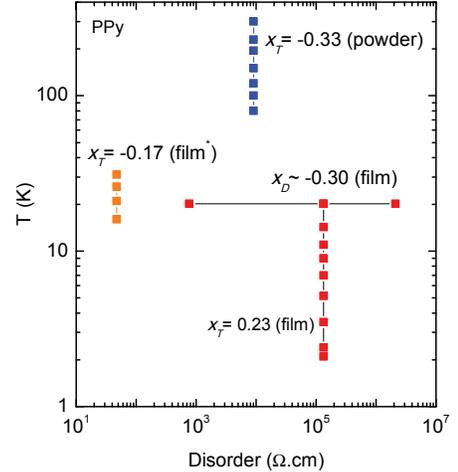}
\caption{(Color online) Measurement paths along with associated nonlinearity exponents in various doped polypyrrole samples in Temperature-Disorder plane. Disorder here is parameterized by the Ohmic resistivity at 20 K. The data marked with star are from Ref. \onlinecite {ribo98}. Note that data from a sample simultaneously belongs to two intersecting lines.}
\label{fig.13}
\end{figure}

\subsection{Non-linearity exponents $x_M$}

Non-linearity exponents are the concrete outcome of the adopted scaling procedure and are displayed in Table III for various CP systems under different conditions. An exponent can be obtained in two ways. One method has been already illustrated in Figs. 3, 5, 7, 8 and 9. This involves collapsing \textit{I-V} curves gathered at different values of some parameter \textit{M} (e.g. temperature) by suitable choices of $F_o$ and $\sigma_o$. The exponent is then obtained using Eq. (\ref{eq:fscale}). This method does not require any knowledge of the scaling function and is solely based upon the scaling property of the \textit{I-V} curves. The other method makes use of the explicit functional form of a \textit{I-V} curve (or, scaling function), when available. It has been shown in the above that GM-expressions (Eqs. \ref{eq:gms} and \ref{eq:scgms}) describe the relevant data in an excellent manner. Therefore, we can use the expression (Eq. \ref{eq:ne2}) for obtaining $x_M$ in terms of model parameters, both direct and derived. $y_n$'s are derived from plots like those shown in Fig. 12. There are as many $y_n$'s as the number of inelastic tunneling channels. According to Eq. (\ref{eq:ne2}), each one of them should yield the same $x_T$ as an yet another test of consistency in applicability of the GM-model to CP's. This seems to be well borne out within errors (Table III) when the contribution from the second channel is large enough to yield reliable values as in PPy (powder) and PPy (film) (this work). Considering the fact that digitization errors are not accounted in the values quoted in case of samples of other works, agreement between values of the exponents obtained using two methods is quite reasonable. In case of measurements at different magnetic fields, since conductivities are related by some constant factors, we have $y_n=1$ for all \textit{n}. Eq. (\ref{eq:ne2}) leads to $x_B=0$ in agreement with the value from scaling consideration as discussed earlier. It may be noted that $y_n$'s in Table III for both positive and negative exponents satisfy inequalities of Eq. (\ref{eq:y1yn}). Also, all determined $x_M$'s lie between the bounds -1/2 and 3/4 (see the point \textit{i}). Incidentally, putting Eq. (\ref{eq:ynexp}) into Eq. (\ref{eq:ne2}) yields $x_M = 1/(n+2)$ instead of a constant value.

Clearly, there is no universal exponent for CP's. There is no universal exponent even for a given conducting polymer. As seen in Table III, the same PPy sample exhibits two different values of the exponent, 0.23 and 0 depending upon the variable \textit{M}. Three different PPy samples exhibits as many values of the exponent $x_T$ (-0.33, 0.23 and -0.17), both positive and negative. This plethora of values could simply indicate that $x_M$ depends upon the path of measurement in the variable space that also includes quenched disorder. Fig. 13 illustrates four such paths of measurements in PPy samples along with the corresponding exponents in Temperature-Disorder plane. Obviously, unraveling of the details of such dependency in the variable space has to await theoretical understanding of $x_M$ that is lacking at present. In fact, different values of $x_M$ in the same system prove that the hopping network is in both physical as well as energy space in contrast to the conduction network in composites at room temperature\cite {stauffer}, that is supposed to be purely geometrical in real space. The field scale can not be simply determined by geometrical topology of disorder alone. Otherwise, temperature that only provides an energy scale could not have an effect on the scale. This is consistent with result, $x_B=0$ since magnetic fields do not change either energy levels of charge carriers or disorder, and therefore, do not warrant a change in the field scale.

\begin{figure}
\includegraphics[width=6cm]{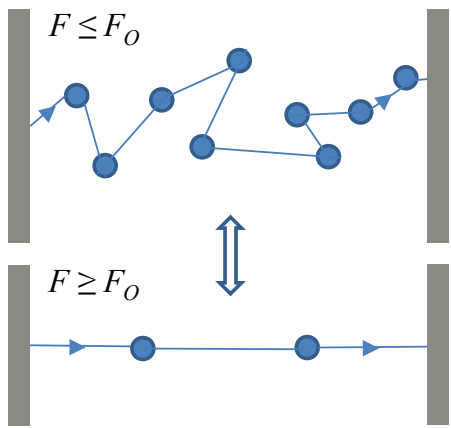}
\caption{(Color online) Schematic diagram of possible hopping paths in real space between electrodes for fields (\textit{F}) less (upper) and more (lower), than an onset field ($F_o$). The solid circles represent localized states. The lower path corresponds to two-step tunneling, the lowest non-Ohmic channel according to Glatzman-Matveev (Ref. \onlinecite{matveev88}). See text for further details.}
\label{fig.14}
\end{figure}

\subsection{Final picture and Issues with the GM-model}
It is evident that the field-dependent data in CP's are in excellent agreement with the GM-expression (Eq. \ref{eq:gms}) and the modified one (Eq. \ref{eq:scgms}). However, in spite of this unexpected success the applicability of the GM-model remains a moot issue. Let us consider the broad picture of transport in lightly doped CP's, that emerges from the results presented here and its analysis. The phenomena of scaling and the GM-expression as the scaling function appears to hold good irrespective of the zero-bias conduction mechanism. Except for PA nanofibres, all other systems considered in this work exhibit VRH-type transport at low bias. In contrast, the model considering conduction across a barrier of width \textit{w} predicts that the temperature dependence of conductance in the limit $eV \ll k_B T$ is given by a temperature-independent term representing direct tunneling plus a series of terms that are sequentially triggered as the temperature is increased. The series is similar to Eq. (\ref{eq:gms}) with $k_B T$ replacing \textit{eV} with the lowest one being proportional to $T^{4/3}$. Consider now the problem of non-Ohmic conduction at a low temperature ($eV \gg k_B T$). In the model, the conduction across a barrier proceeds from direct tunneling to resonant tunneling to directed hopping along quasi-one dimensional paths of localized states and finally, in the bulk limit ($w \gg R_h$) to variable range hopping\cite {beasley95}. Thus, the VRH that appears in the original GM-model as a limiting process appears in the present problem at the very outset of low bias. So, how does one envisage the evolution of a percolative\cite {efrosbook} trajectory of a charge carrier in VRH (Ohmic) regime as shown schematically in the upper panel in Fig. 14 to a linear trajectory in the non-Ohmic regime as shown in the lower panel of Fig. 14? As the applied field is increased beyond the onset value, the conduction becomes non-Ohmic. According to the GM-picture, non-Ohmic part in conductance results from progressive triggering of the inelastic multi-step tunneling channels starting from the lowest two-step tunneling (at least when the latter has a non-zero contribution) shown in the lower panel of the figure. The conductance corresponding to two-step tunneling is proportional to $\exp (-w/3a)$ where \textit{w} is the barrier width. Obviously, it can not be a macroscopic length ($w \gg a$) typical of the samples considered here. If we take \textit{w} to be of the order of disordered region along a polymer fibril, then one has to consider many such processes throughout a sample in comparison of only one in the original model. If VRH applies to the whole sample it is not clear what will be the total trajectory corresponding to two-step tunneling. Furthermore, since Ohmic-non-Ohmic transition is a continuous process it is hard to imagine a transformation of paths of a charge carrier as indicated in the figure. To summarize, one has to reckon with the fact that the expression (Eq. \ref{eq:gms}) that was supposed to describe hopping across a thin film only is now required to describe hopping across apparently multitudes of such 'films' embedded inside bulk systems.

Apart from the conceptual problems outlined above, we list here the interesting features involving particularly the coefficients $\sigma_n$ that were observed and need to be explained theoretically. Firstly, the incidence of channel selection. This is not a universal phenomenon. It occurs in CP's and probably, other materials but is absent in tunnel junctions\cite {beasley95}. It is necessary to understand the exact conditions in the GM-model that give allow such a selection phenomenon, and then how those conditions may be satisfied in CP's. Secondly, the assumption (Eq. \ref{eq:gmcoeff}) that makes scaling analysis possible within the GM-model. Thirdly, Eq. (\ref{eq:yn}) that relates each channel coefficient to the linear conductivity. Simple arguments yield only $y_n < 1$ for $n>1$ (Eq. \ref{eq:ynexp}). But $y_n > 1$ is necessary to have negative nonlinearity exponents.

\section{Conclusion}

In this paper, we have reported field-dependent conductivities in various CP systems as a function of temperature and quenched disorder. We demonstrated that each of various conducting polymer systems possess a \textit{single} field scale and exhibits the associated scaling. A phenomenological scaling equation that led to extraction of nonlinear exponents was used to analyze the nonlinear transport data. It was argued that experimental evidence points to the linear conductivity as a legitimate scaling variable. Surprisingly, the GM-expression for multi-step tunneling proves to be an excellent fit to the \textit{I-V} curves as well as the scaled curves. A couple of assumptions are made to make the GM-model compatible with scaling. Experimental values of the exponents fall within the predicted limits of -0.5 and 0.75. A theory capable of explaining the nonlinear exponents, particularly negative ones, is lacking. The value of the exponent depends upon the path of measurement in the variable space. Some issues concerning applicability of the GM-model to CP's have been discussed.

The scaling that has been observed in CP's has been also found in many other disordered systems including composites. It is believed that such scaling may be truly an universal feature of disordered systems particularly with localized states. All the samples considered here are three-dimensional. It will be interesting to know how such scaling fare in lower dimensions.

\section{Acknowledgments}
This work was supported  in part by the Department of Science and Technology, Government of India through Major Research Project No: SR/S2/CMP-0054/2008.

\bibliography{nonohmic}

\begin{thebibliography}{10}
\bibitem{paper25_1}%
  \BibitemOpen
  \bibinfo {note} {The exponent is 1.75 according to Ref. \onlinecite
  {aleshin04}.}%
  \bibAnnoteFile{Stop}{paper25_1}%
\end{thebibliography}

\end{document}